\documentclass[10pt,preprint]{aastex}

\def\a0{a_0}
\def\msun{M_{\odot}}
\def\kms{{\rm km~ s}^{-1}~}

\def\deg{^o}
\def\kpc{{\rm kpc~}}
\def\mpc{{\rm Mpc~}}

\def\deg{^o}

\def\beq{\begin{equation}}
\def\eeqno#1{\label{#1}\end{equation}}

\def\msun{M_{\odot}}
\def\az{a_{0}}
\def\mz{\mu_{0}}
\def\l0{\ell_{0}}

\def\rar{\rightarrow}

\def\r{\rho}

\def\drt{d^3r}
\def\f{\phi}
\def\grad{\vec\nabla}
\def\div{\vec \nabla\cdot}
\def\gf{\grad\phi}

\def\vr{\textbf{r}}

\def\vv{\textbf{v}}
\def\vV{\textbf{V}}
\def\va{\textbf{a}}
\def\vq{\textbf{q}}
\def\vQ{\textbf{Q}}
\def\vF{\textbf{F}}
\def\vA{\textbf{A}}
\def\azun{\times10^{-8}{\rm cm~s^{-2}}}
\def\hubun{~{\rm km ~s}^{-1} {\rm Mpc}^{-1}}
\def\mpc{~{\rm Mpc}}
\def\kpc{~{\rm kpc}}

\def\cm{{~\rm cm}}
\def\kms{~{\rm km~s^{-1}}}
\def\hubble#1{H_0=#1\hubun}
\def\l{\lambda}

\usepackage{natbib}
\usepackage{amssymb}
\usepackage{epsfig}
\begin{document}

\title{The MOND paradigm}

\author{Mordehai Milgrom}
\affil{Center for Astrophysics, Weizmann Institute, Rehovot 76100,
Israel}

\begin{abstract}{I review briefly different aspects of the MOND
paradigm, with emphasis on phenomenology, epitomized here by many
MOND laws of galactic motion--analogous to Kepler's laws of
planetary motion. I then comment on the possible roots of MOND in
cosmology, possibly the deepest and most far reaching aspect of
MOND. This is followed by a succinct account of existing underlying
theories. I also reflect on the implications of MOND's successes for
the dark matter (DM) paradigm: MOND predictions imply that baryons
alone accurately determine the full field of each and every
individual galactic object. This conflicts with the expectations in
the DM paradigm because of the haphazard formation and evolution of
galactic objects and the very different influences that baryons and
DM are subject to during the evolution, as evidenced, e.g., by the
very small baryon-to-DM fraction in galaxies (compared with the
cosmic value). All this should disabuse DM advocates of the thought
that DM will someday be able to reproduce MOND: it is inconceivable
that the modicum of baryons left over in galaxies can be made to
determine everything if a much heavier DM component is present.}

\end{abstract}
\section{Introduction: the MOND paradigm}
\label{section1} MOND is an alternative paradigm to Newtonian
dynamics, whose original motivation was to explain the mass
discrepancies in galactic systems without invoking dark matter (DM)
(Milgrom 1983a). It constitutes a modification of dynamics in the
limit of low accelerations that rests on the following basic
assumptions: (i) There appears in physics a new constant, $\az$,
with the dimensions of acceleration. (ii) Taking the formal limit
$\az\rar 0$ in all the equations of physics restores the equations
of classical (pre-MOND) dynamics. (iii) For purely gravitational
systems, the opposite, deep-MOND limit, $\az\rar \infty$, gives
limiting equations of motion that can be written in a form where the
constants $\az$ and $G$, and all masses in the problem, $m_i$,
appear only in the product $m_iG\az=m_i/\mz$, where
$\mz\equiv(G\az)^{-1}$ (Milgrom 2005)\footnote{By this I mean that
starting with equations that involve (including in derivatives)
$\vr$, $t$, $G$, $\az$, $m_i$, and gravitational field degrees of
freedom, we can rewrite them, possibly by redefining the
gravitational field, so that only $\vr$, $t$, $m_i/\mz$, and the
gravitational field appear.}. This last fiat reproduces the desired
MOND phenomenology for purely gravitational systems. A MOND theory
is one that incorporates the above tenets in the nonrelativistic
regime.
\par
Since all our knowledge of MOND comes, at present, from the study
of purely gravitational systems (galactic systems, the solar
system, etc.) it is still an open question how exactly to extend
the third MOND tenet to systems involving arbitrary interactions.
One possibility is to require that for $\az\rar\infty$, the
limiting equations of motion can be brought to a form where $G$,
$\az$, and $m_i$ appear only as $G\az^2=\az/\mz$ and $m_i/\az$.
This requirement will automatically cause $\az$, $G$, and $m_i$ to
appear as $m_i/\mz$ in the deep MOND limit of purely gravitational
systems, as in such cases $G$ and $m_i$ always appear in equations
as $Gm_i$. Such a general requirement would also replace Newton's
second law $\vF=m\va$ by $\vF=m\vQ/\az$ in the deep MOND regime,
where $\vQ$ is some functional with dimensions of acceleration
squared that does not depend on $\az$.
\par
Detailed reviews of phenomenological and theoretical aspects of the
paradigm can be found, for instance, in Sanders and McGaugh (2002),
and, more recently, in Scarpa (2006) and in Bekenstein (2006).

It follows from the third tenet that an underlying MOND theory must
be nonlinear in the sense that an acceleration of a test particle
due to a combination of several fields is not simply the sum of the
accelerations produced by the individual fields. Take, as an
example, the purely gravitational case where we modify the equations
for the gravitational field. Linearity would mean that in the
nonrelativistic limit of the theory the acceleration of a test
particle at position $\vr$ in the field of $N$ masses $m_i$ at
positions $\vr_i$ is given by
 \beq \va=\sum_{i=1}^N m_i q_i(G, \az, \vr, \vr_1,...,\vr_N). \eeqno{lino}
 The third assumption says that in the deep MOND case $q_i\propto
\mz^{-1}$, but this is dimensionally impossible. Clearly, the
acceleration produced even by a single point  cannot be linear in
its mass, as we shall see explicitly below.
\par
It also follows from the third tenet, and the assumption that $\az$
is the only new dimensioned constant, that the deep MOND limit of
any theory must satisfy the following scaling laws for purely
gravitational systems: In general, on dimensional grounds, all
physics must remain the same under a change of units of length
$\ell\rar\lambda\ell$, of time $t\rar\lambda t$, and no change in
mass unit $m\rar m$. Under these we have $\az\rar \lambda^{-1}\az$
and $G\rar \lambda G$; so, the constant $\mz$, which alone appears
in the limiting theory, is invariant under the
scaling\footnote{Other dimensioned quantities, such as the
gravitational potential, have to be scaled appropriately. Note that
velocities are invariant.}. This tells us that we are, in fact,
exempt from scaling the constants of the theory when we scale
$(\vr,t)\rar \lambda(\vr,t)$. The theory is thus invariant under
this scaling; namely, if a certain configuration is a solution of
the equations, so is the scaled configuration. Specific theories may
have even higher symmetries; for example, the above scaling property
is only part of the conformal invariance of the deep MOND limit for
the particular MOND theory of Bekenstein \& Milgrom (1984), as found
in Milgrom (1997).

\par
As a corollary of the scaling invariance we have: if $\vr(t)$ is a
trajectory of a point body in a configuration of masses $m_i$ at
positions $\vr_i(t)$ (which can be taken as fixed, for example),
then $\hat\vr(t)=\lambda\vr(t/\lambda)$ is a trajectory for the
configuration where $m_i$ are at $\lambda\vr_i(t/\lambda)$, and
the velocities on that trajectory are $\hat\vV(t)=\vV(t/\lambda)$.
(An extended mass changes its size and density such that the total
mass remains the same. A point mass remains a point mass of the
same value.)
\par
Since $\mz$ has dimensions of $mt^4/\ell^4$, another scaling under
which it is invariant is $m_i\rar\lambda m_i$, $\vr_i\rar\vr_i$,
$t\rar\lambda^{-1/4} t$. (The most general scaling allowed is
$m\rar\lambda m$, $\vr_i\rar\kappa\vr_i$,
$t\rar\kappa\lambda^{-1/4}t$.) This means that for a purely
gravitational system deep in the MOND regime, scaling all the masses
leaves all trajectory paths the same but the bodies traverse them
with all velocities scaling as $m^{1/4}$; accelerations then scale
as $m^{1/2}$.

\par
It is enlightening to draw an analogy between the role of $\az$ in
MOND with the role of $\hbar$ in quantum physics, or that of $c$ in
relativity. These constants, each in its own realm, mark the
boundary between the classical and modified regimes; so formally
pushing these boundaries to the appropriate limits ($c\rar\infty$,
$\hbar\rar0$, $\az\rar 0$) one restores the corresponding classical
theory (for quantum theory, in some weak sense). In addition, these
constants enter strongly the physics in the modified regime where
they feature in various phenomenological relations. For example,
$\hbar$ appears in the black body spectrum, the photoelectric
effect, atomic spectra, and in the quantum Hall effect. The speed of
light appears in the Doppler effect, in the mass vs. velocity
relation, and in the radius of the Schwarzschild horizon. Without
the respective, underlying theory, these disparate phenomena would
appear totally unrelated, and the appearance of the same numerical
constant in all of them would constitute a great mystery. The MOND
paradigm similarly predicts a number of laws related to galactic
motion, some of which are qualitative, but many of which are
quantitative and involve $\az$. Since they appear to be obeyed by
nature it should indeed be a great mystery why that should be so
without the underlying MOND paradigm (i.e. with Newtonian dynamics
plus DM). I now discuss some of these predictions in some detail.

\section{MOND laws of galactic dynamics}
\label{section2}
 A MOND theory based on the above basic premises
should predict everything about the acceleration field of an object,
such as a galaxy, based on the baryon distribution alone. But even
before looking at detailed predictions for individual systems, it is
possible and useful to distil a number of corollaries that follow
essentially from the basic premises themselves. This helps focus
attention on some unifying, general laws, which may be likened to
Kepler's laws of planetary motion. MOND, of course, predicts more
then just these laws.

I list some such relations below and explain how they come about. It
is important to realize that some of these actually contradict the
predictions of CDM, and that those that do not are independent in
the context of DM. They are so in the sense that one can construct
galaxy models with baryons and DM that satisfy any set of these
predictions but not any of the others. So, in the context of DM each
would require a separate explanation. The MOND predictions
concerning the mass discrepancies in galactic systems depend only on
the present day baryon distribution. In contrast, the expected
discrepancies; i.e., the relative quantities and distributions of
baryons and DM in such systems depend strongly on their unknown (and
unknowable) formation history, a point which I expand on in section
\ref{section9}.

In deriving the predictions below I assume that the theory
involves only $\az$ as a new dimensioned constant.
\begin{itemize}

\item[1.] The orbital speed around an arbitrary isolated mass
becomes independent of the radius of the orbit in the limit of very
large radii. This means, in particular, that the rotational velocity
on a circular orbit becomes independent of the orbital radius at
large radii--asymptotic flatness of rotation curves: $V(r)\rar
V_{\infty}$. This quantitative behavior was the cornerstone on which
MOND was built; so it was introduced by hand, based on anecdotal
evidence existing at the time.  But once taken as axiom of MOND it
has become a binding prediction. It follows straightforwardly from
the above length-time scaling property of the deep MOND
limit\footnote{Just as the analogue third Kepler law for Newtonian
dynamics, $V(r)\propto r^{-1/2},$ follows from the fact that masses
and G always appear in the combination $MG$, which is invariant
under $\vr\rar\lambda\vr$, $t\rar \lambda ^{3/2}t$, under which
$V\rar \lambda^{-1/2}V$.}; but it is worthwhile seeing an explicit
derivation: On dimensional grounds, the acceleration of a test
particle on an arbitrary orbit around a point mass, $M$, when it is
at a distance $r$ on the orbit, has to scale with $M$ and $r$ as
 \beq a\sim {MG\over r^2}\nu\left(\vq,{MG\over
r^2\az}\right), \eeqno{i} where $\nu(\vq,y)$ can depend, through
some dimensionless parameters $\vq$, on the geometry of the orbit
and on the position on the orbit\footnote{I assume implicitly that
for a mass distribution of total mass $M$ that is bounded in a
volume of size $R$, all that enters the motion of a test particle at
large radii ($r\gg R$) is $M$, while the exact spatial distribution
of the mass is immaterial. In other words, I assume that the motion
of a test particle in the field of a point mass is independent of
the way the limit of the point mass is taken. This is certainly true
for modified inertia theories in which the gravitational field
equation is the standard one. For modified gravity theories it seems
an obvious assumption, but, in principle, it has to be checked (it
does hold for all the theories studied to date).}. From the basic
premises of MOND we must have:

\beq \nu(\vq,y)\approx\left\{\begin{array}{l@{\quad:\quad} l}  1 &
y\gg 1\\\eta(\vq) y^{-1/2} & y\ll 1\end{array}\right..\eeqno{ii}

 This means that at large $r$ we have $a=\eta(\vq)(MG\az)^{1/2}/r$.
This, in turn, means that the orbital speed is given by

\beq
V(M,r,\vq)=\l(\vq)(MG\az)^{1/4}=\l(\vq)(M/\mz)^{1/4},\eeqno{lul}
 and is thus
invariant to scaling of the orbit. This holds for any MOND theory
(be it modified gravity or modified inertia--see below) and for
any orbit. For circular orbits in an axisymmetric potential $\l$
has to be constant as it cannot depend on orbital phase. The value
of $\az$ is normalized so that for circular orbits $\l=1$.

\par
The function $\nu(\vq,y)$ is one example of the appearance of a so
called interpolating function in MOND [the name is usually reserved
for the function $\mu(x)$ related to $\nu$ by $\mu(x)=i(x)/x$, where
$i(x)$ is the inverse of $y\nu(y)$]. In some formulations of MOND
(as in modified gravity--see below) there is a single interpolating
function that appears in the theory. In other formulations there
isn't a unique one (as, for example here, where $\nu$ can depend on
$\vq$, in general). In a given context it interpolates between the
known behaviors in the deep MOND regime and the classical (pre MOND)
regime.

\item[2.]

MOND predicts a relation between the total mass of a body and the
asymptotic circular velocity around it. It follows from
eq.(\ref{lul}) that $V_{\infty}^4=MG\az$ (Milgrom 1983b). This can
be tested directly for disc galaxies (see confirming analysis in
McGaugh et al. 2000, McGaugh 2005a,b). This predicted
mass-asymptotic-speed relation is in the basis of the traditional,
empirical Tully-Fisher (TF) relation between galaxy luminosity and
rotational speed, from which, however, it differs in essence.
 One finds in the literature various empirical plots that are termed TF
relations, between some luminosity measure and some velocity
measure. Many of these plots are not useful for comparison with
theoretical predictions. MOND dictates exactly what should be
plotted: the total baryonic mass against the rotational velocity on
the flat part of the rotation curve--aka the baryonic Tully-Fisher
(TF) relation. Using, as in the original TF relation, the luminosity
as a mass measure will not do; at best, it represents only the
stellar mass, leaving out the gas mass. This is clearly demonstrated
in Fig.\ref{fig1} (see also Milgrom \& Braun 1988) where a tight
relation is followed only for high mass galaxies, which are
dominated by stellar mass. The same relation holds for lower mass
galaxies, which are gas rich, only if we include the gas mass. Note
also that the use of a velocity measure that is heavily weighted by
smaller radii (e.g. optical velocity measures) artificially distorts
the mass-velocity plot, introducing artificial scattering and
biasing the slope of the relation: For low mass galaxies the inner
velocities are, typically, smaller than the asymptotic ones, while
for high mass galaxies the opposite is true.

\begin{figure}[h]
\begin{tabular}{rl}
\tabularnewline
\includegraphics[width=0.9\columnwidth]{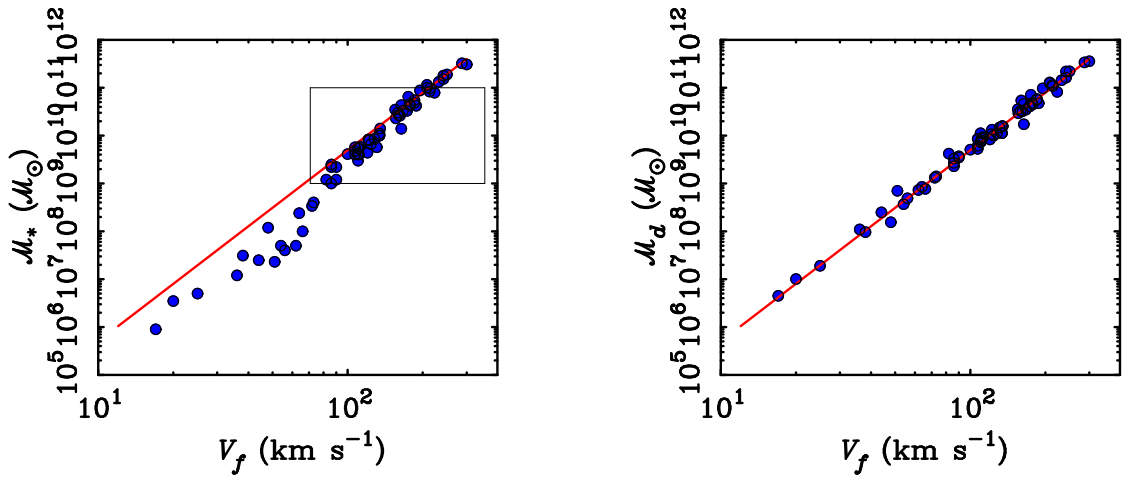} \\
\end{tabular}\par
\caption{Galaxy mass plotted against the rotation curve plateau
velocity. Left: analog of the traditional Tully-Fisher plot with
mass in stars only. Right: The total mass including that of gas. The
solid line has the log-log slope of 4, predicted by MOND, and is not
a fit (McGaugh 2005b) (the small rectangle shows where past analysis
had concentrated).} \label{fig1}
\end{figure}

\item[3.]

In a disc galaxy, whose rotation curve is $V(r)$, that has high
central accelerations ($V^2/r>\az$ in the inner regions), the mass
discrepancy appears always around the radius where $V^2/r=\az$. In
galaxies whose central acceleration is below $\az$ (low surface
brightness galaxies--LSBs) there should appear a discrepancy at all
radii (Milgrom 1983b, and for a confirmation see, e.g., McGaugh
2006).

\item[4.] For a concentrated mass, $M$, well within its transition
radius, $r_t\equiv (MG/\az)^{1/2}$, $r_t$ plays a special role
(somewhat akin to that of the Schwarzschild radius in General
Relativity) since the dynamics changes its behavior as we cross from
smaller to larger radii. For example a shell of phantom DM may
appear around this radius (Milgrom \& Sanders 2008).

\item[5.] Isothermal spheres  have mean surface densities
$\bar\Sigma\lesssim \Sigma_0\equiv \az/G$ (Milgrom 1984) underlying
the observed Fish law for quasi-isothermal stellar systems such as
elliptical galaxies (see discussion in Sanders and McGaugh 2002).
This follows from the fact that Newtonian, self gravitating,
isothermal spheres (ISs) have an enclosed mass that increases
linearly with radius, and they thus have an infinite total mass. The
only way to avoid this is via a mass cutoff provided by MOND, which
can only be felt around $r_t$. But, by definition, the mean surface
density within $r_t$ is $\sim\Sigma_0$. Unlike Newtonian ISs, MOND
ISs have, indeed, a finite total mass. They can have arbitrarily
small mean surface densities.

\item[6.]   For spheroidal systems a mass-velocity-dispersion relation $\sigma^4\sim
MG\az$ is predicted under some circumstances. According to MOND,
this is the fact underlying the observed Faber-Jackson relation for
elliptical galaxies, which are approximately isothermal spheres
(Milgrom 1984). For instance, this relation holds approximately for
all isothermal spheres having a constant velocity dispersion
$\sigma$ and constant velocity anisotropy ratio $\beta$ (Milgrom
1984). Such spheres are also characterized by a mass $M$, and some
mean radius, $R$. On dimensional grounds we have, in a given MOND
theory: $\sigma^4=f\left (\beta,{MG\over R^2\az}\right)MG\az$. From
the basic premises of MOND, when $MG/R^2\az$ is small compared with
1, $f$ becomes independent of this quantity. Prediction 5 says that
this ratio cannot be large compared with 1. These give $\sigma^4\sim
\hat f(\beta)MG\az$ ($\hat f$ depends weakly on $MG/R\az$). (See
Milgrom 1984 for detailed calculations in a particular MOND theory).
In the same MOND theory Milgrom (1994b) derived a general relation
for arbitrary, stationary, self gravitating, many-particle systems
(not necessarily isothermal, or spheroidal) in the deep MOND regime,
of the form $\sigma^4=(4/9)MG\az$, where $\sigma$ is the 3-D rms
velocity dispersion. Sanders (2000) discussed a generalization of
isothermal spheres that better fit observed ellipticals.

Note the difference between this relation and the mass-velocity
relation in law 2. The latter holds for all systems and is exact,
but involves the large radius asymptotic rotational speed; the
former is limited to either isothermal spheres or to low
acceleration systems (e.g., it does not apply to stars, which are
neither), is only approximate even for these, and involves the bulk
velocities.

\item[7.]
There is a difference in the dynamics, and hence in the stability
properties, of discs with mean surface density $\bar\Sigma\lesssim
\az/G$ and $\az/G\lesssim\bar\Sigma$ (Milgrom 1989a, Brada \&
Milgrom 1999b, Tiret \& Combes 2007a,b).

\item [8.] The excess acceleration that MOND produces over
Newtonian dynamics, for a given mass distribution, cannot much
exceed $\az$ (Brada \& Milgrom 1999a). This simply follows from the
fact that MOND differs from Newtonian dynamics only when the
accelerations are around or below $\az$. Put in terms of DM this
MOND prediction would imply that the acceleration produced by a DM
halo alone can never much exceed $\az$, according to MOND. There is
no known reason for this to hold in the context of the DM paradigm.
This prediction was confirmed by Milgrom \& Sanders (2005) for a
sample of disc galaxies.

\item[9.]
An external acceleration field, $g_e$, enters the internal dynamics
of a system imbedded in it. For example, if the system's intrinsic
acceleration is smaller than $g_e$ , and both are smaller than
$\az$, the internal dynamics is quasi-Newtonian with an effective
gravitational constant $G\az/g_e$ (Milgrom 1983a, 1986a, Bekenstein
\& Milgrom 1984). This was applied to various astrophysical systems
such as dwarf spheroidal galaxies in the field of a mother galaxy,
warp induction by a companion, escape speed from a galaxy, departure
from asymptotic flatness of the rotation curve, and others (see,
e.g., Brada \& Milgrom 2000a,b, Famaey, Bruneton, \& Zhao 2007,
Angus \& McGaugh 2007, and Wu et al. 2007). This external field
effect (EFE) follows from the inherent nonlinearity of MOND. It
appears in somewhat different ways in all versions of MOND studies
to date, but I am not sure that it is a general consequence of the
basic premises alone.
\par
In some applications the external field effect can be mimicked by a
cutoff in a DM halo; but, others of its consequences violate the
strong equivalence principle, and are thus not reproducible with DM;
e.g., the time dependent, non-tidal effects on the structure and
internal kinematic of dwarf spheroidals that fall in the field of a
mother galaxy, and the induction of warps discussed by Brada \&
Milgrom (2000a,b).

\item[10.] The nonlinearity of MOND also leads to a breakdown of the thin
lens approximation: Two  different mass distributions having the
same projected surface density distributions on the sky, do not
produce the same lensing effect as is the case, approximately, in
General Relativity (Mortlock and Turner 2001, Milgrom 2002a). For
example, consider a chain of $N$ equal, point masses $m$ far apart
from each other along the line of sight [much farther than their
individual transition radius $(mG/\az)^{1/2}$], but closer together
than the observer-lens and the lens-source distances. In standard
dynamics they act as a single point mass $Nm$. In deep MOND the
gravitational field scales as $m^{1/2}$, so the effect for a single
mass $Nm$ scales as $(Nm)^{1/2}$, while that of the chain scales as
$Nm^{1/2}$ (Milgrom 2002a). Some implications and applications of
this are discussed, e.g., in Milgrom \& Sanders (2008), and in Xu et
al. (2007). This MOND law conflicts with the predictions of DM.

\item[11.] Disc galaxies are predicted to exhibit a disc mass
discrepancy, as well as the spheroidal one that is found for any
mass. In other words, when MOND is interpreted as DM we should
deduce a disc component of DM as well as a spheroidal one (Milgrom
1983b,2001). The reason: the dynamical surface density in a disc is
deduced from the normal component of acceleration just outside it.
Since the MOND prediction for this acceleration differs from the
Newtonian value (and is generically larger) MOND predicts that
Newtonian analysis will find a higher dynamical surface density than
is observed. Unlike the spheroidal component, which extends to large
radii, the disc component is confined to the baryonic disc: Where
there is no baryonic disc there is no jump in the normal component
of the acceleration and hence no mass discrepancy is predicted
there. In LSB galaxies this phantom disc should be found everywhere
in the baryonic disc; in HSB galaxies only where $V^2/r\lesssim\az$.
For thin discs, at radii where the radial acceleration dominates the
perpendicular one, and $V^2/r\ll\az$, the surface density of the
phantom disc is $\sim (V^2/r\az)^{-1}$ times the baryon surface
density in the disc. For results of analysis in support of this
prediction see Milgrom (2001), Kalberla et al.
(2007)\footnote{Interestingly, Kalberla et al. find that in order to
reproduce the observed flaring data of the Milky Way disc one needs,
in addition to a spheroidal halo of DM, a disc component and also a
disc-like ring of DM centered at about $16 \kpc$. I find in Milgrom
(2008b) that for most forms of $\mu(x)$, the fictitious DM disc
predicted by MOND has a maximum in its surface density at a radius
of the order of the transition radius, and could thus mimic a
disc-plus-ring component of DM as is found in the analysis of
Kalberla et al.. This disc-like ring predicted by MOND is analogous
to the phantom shell of ``DM'' predicted for concentrated masses, as
discussed in Milgrom \& Sanders (2008) (see law number 4 above). I
show some such predicted phantom discs in Fig. \ref{figx} for a
heuristic model of the Milky Way. The model consists of a thin
exponential disc of scale length 0.3 in units of the transition
radius, and a de Vaucouleur sphere with effective radius 0.1 in
these units. The disc-to-bulge mass ratio is 1:0.3.  (For a MW mass
of $10^{11}\msun$, or asymptotic rotational speed of $200\kms$, the
transition radius is at $\approx 11\kpc$, assuming $\az=1.2\azun$.)
S\'anchez-Salcedo et al. use interpolating functions of the forms
$\mu_1$ and $\mu_2$ (in the nomenclature of Milgrom \& Sanders
2008), which as we see in Fig. \ref{figx} are not completely in line
with the deductions of Kalberla et al., as these forms predict no
surface density peak or one at smaller radii. Indeed the MOND
results of S\'anchez-Salcedo et al., while satisfactory, are
somewhat lacking, and may do better for other choices of $\mu$ that
give a surface density peak at larger radii.}, and
S\'anchez-Salcedo, Saha, \& Narayan (2007).

This prediction contradicts the expectations for Cold dark matter,
which, being dissipationless, forms only spheroidal halos, not
discs.

\begin{figure}[h]
\begin{tabular}{rl}
\tabularnewline
\includegraphics[width=0.8\columnwidth]{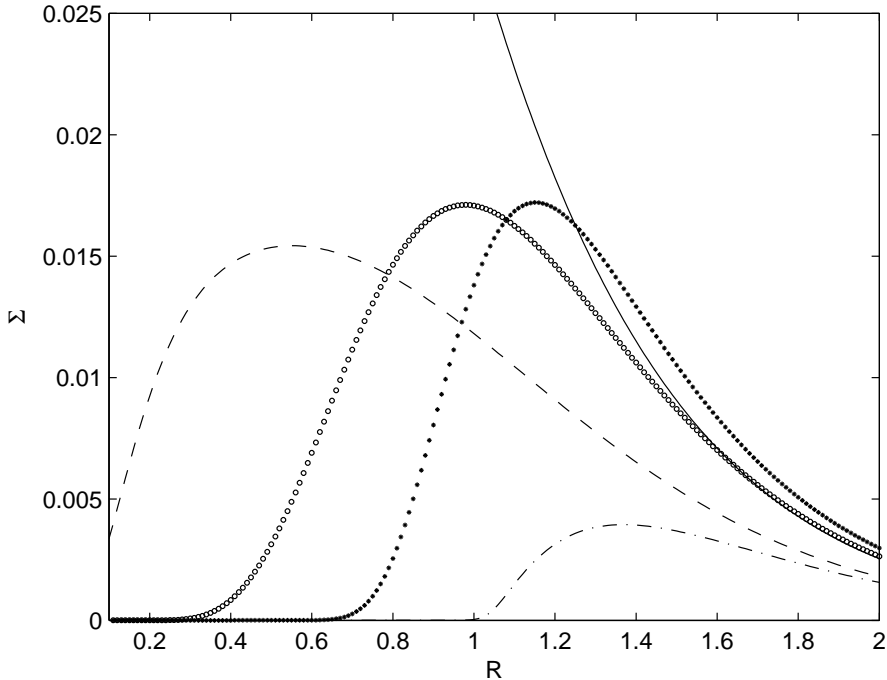} \\
\end{tabular}\par
\caption{The MOND prediction of the fictitious DM disc surface
density in units of $\Sigma_0=\az/G$, as a function of
galactocentric radius in units of the transition radius, for a
heuristic model of the Milky Way. Results for five interpolating
functions are shown using the designation in Milgrom \& Sanders
(2008): $\mu_1$ (solid), $\mu_2$ (dashed), $\bar\mu_{1.5}$
(circles), $\bar\mu_{3}$ (stars), and $\mu_{40}$ (dot-dash).}
\label{figx}
\end{figure}

\item[12.] A DM interpretation of MOND should give negative density
of ``dark matter'' in some locations (Milgrom 1986b): The reason:
there is no guarantee that the divergence of the MOND acceleration
field is larger than $4\pi G\rho$ everywhere, where $\rho$ is the
baryon density. The density distribution deduced from the Poisson
equation may thus fall below the observed baryon source density, in
places. If verified, this would directly conflict with DM.

\end{itemize}

\section{Galaxy rotation curves in MOND}
\label{section4}

The quintessential MOND result, above and beyond the preceding
general laws, is  the  prediction of the rotation curves (RC) of
individual disc galaxy, based only on the observed (baryonic) mass
distribution. The force of this prediction was clear from the
beginning; but, interestingly, it took some five years after the
advent of MOND for the first RC analysis to be performed as
described in Kent (1987) and the amending sequel by Milgrom (1988).
Meaningful analysis had had to await the appearance of extended RCs
afforded by HI observations. Many such analyses followed (Milgrom \&
Braun 1988, Lake 1989 and its rebuttal in Milgrom 1991, Begeman
Broeils \& Sanders 1991, Broeils 1992, Morishima \& Saio 1995,
Sanders 1996, Moriondo Giovanardi \& Hunt 1998, Sanders and
Verheijen 1998, de Blok \& McGaugh 1998, Bottema et al. 2002, Begum
\& Chengalur 2004, Gentile et al. 2004, 2007a, 2007b, Corbelli \&
Salucci 2007, Milgrom \& Sanders 2007, Barnes Kosowsky \& Sellwood
2007, Sanders \& Noordermeer 2007, Milgrom 2007). In all these
analyses one has used the MOND relation

\beq\mu(V^2/r\az)V^2/r=g_N, \eeqno{alge} where $\mu(x)$ is a MOND
interpolating function, and $g_N$ is the Newtonian acceleration
calculated from the observed baryon distribution. Such a relation is
an exact result in modified inertia theories for circular motion in
axisymmetric potentials, as applies here, and is approximate for
modified gravity theories.

Most RC analyses in the past have involved medium- to
low-acceleration galaxies, for which the MOND prediction is not
sensitive to the exact form of the interpolating function, as long
as it satisfies the required small-arguments limit (see e.g.,
Milgrom \& Sanders 2008). Analyses of HSB galaxies, which can
constrain $\mu(x)$, have started in earnest only recently; for
example in Famaey \& Binney (2005), Zhao \& Famaey (2006), Sanders
\& Noordermeer (2007), and Milgrom \& Sanders (2008).

\begin{center}
\begin{figure*}[h]
\begin{tabular}{rl}
\tabularnewline
\includegraphics[width=0.28\columnwidth]{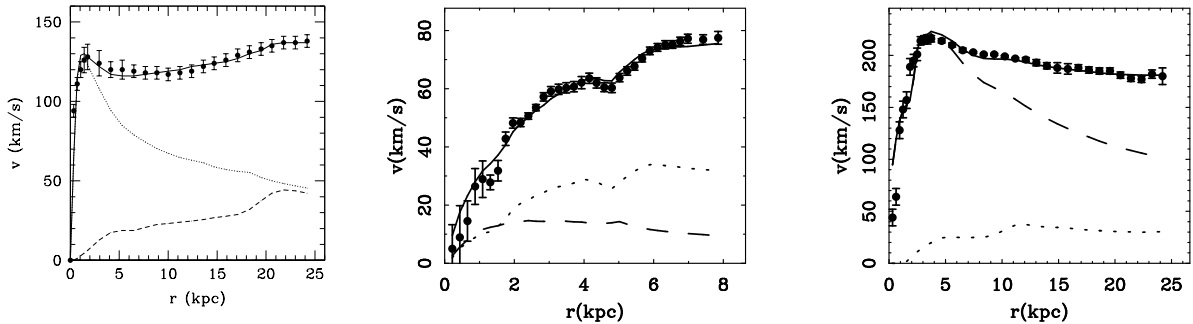} &
\includegraphics[width=0.65\columnwidth]{twogalII}\\
\end{tabular}\par
\caption{The observed and MOND rotation curves (in solid lines) for
NGC 3657 (left), NGC 1560 (cnter),  and NGC 2903 (right). The first
from Sanders (2006a), the last two from Sanders and McGaugh (2002).
Dotted and dashed lines are the Newtonian curves calculated for the
different baryonic components (they add in quadrature to give the
full Newtonian curve). For NGC 3657 the dotted line is for stars and
dashed for the gas, with the reverse for the other two galaxies. }
\label{fig2}
\end{figure*}
\end{center}

\begin{center}
\begin{figure*}[h]
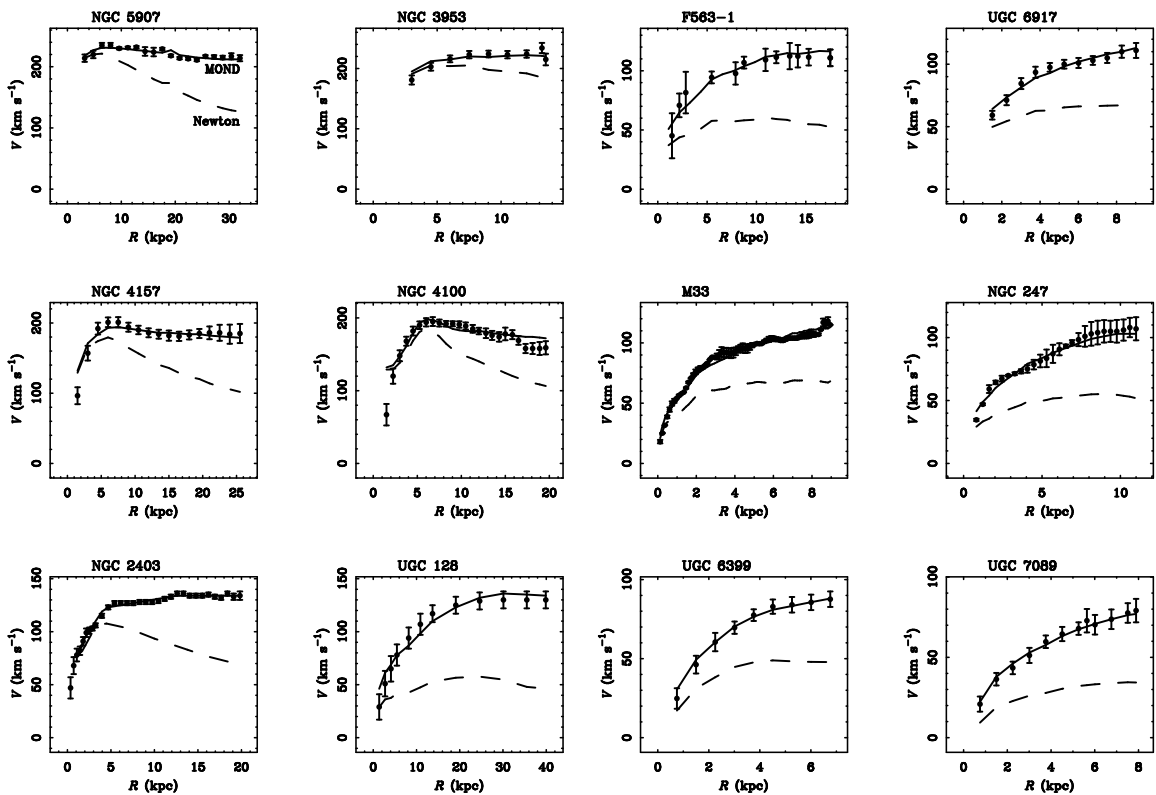

\begin{tabular}{rl}
\tabularnewline
\includegraphics[width=0.45\columnwidth]{rc1} &
\includegraphics[width=0.45\columnwidth]{rc2}\\
\end{tabular}\par
\caption{Additional MOND rotation curves  from McGaugh (private
communication). Dashed lines are the Newtonian curves, the solid
lines the MOND curves.} \label{fig3}
\end{figure*}
\end{center}

\begin{center}
\begin{figure*}[h]
\begin{tabular}{rl}
\tabularnewline
\includegraphics[width=0.45\columnwidth]{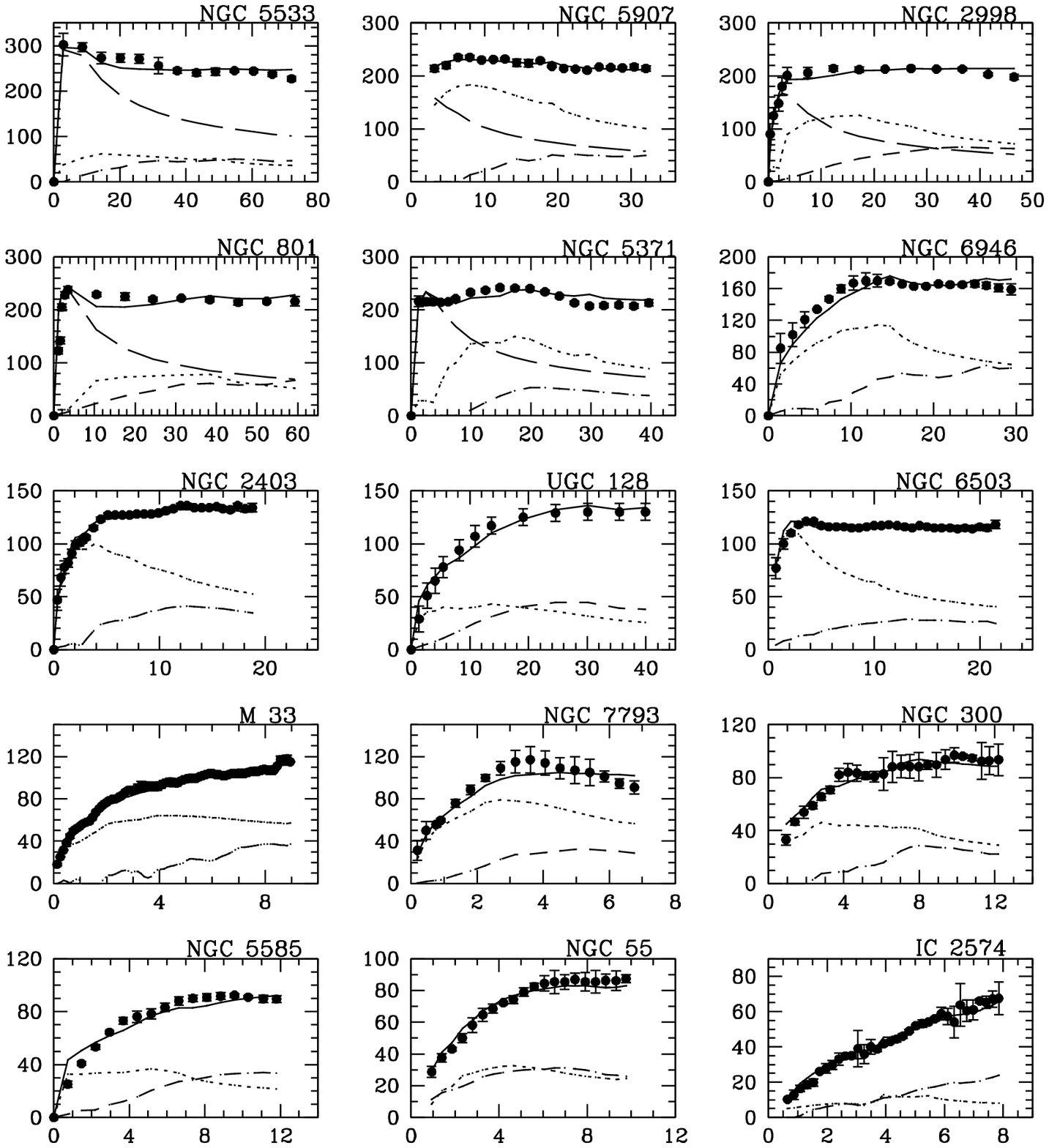} &
\includegraphics[width=0.45\columnwidth]{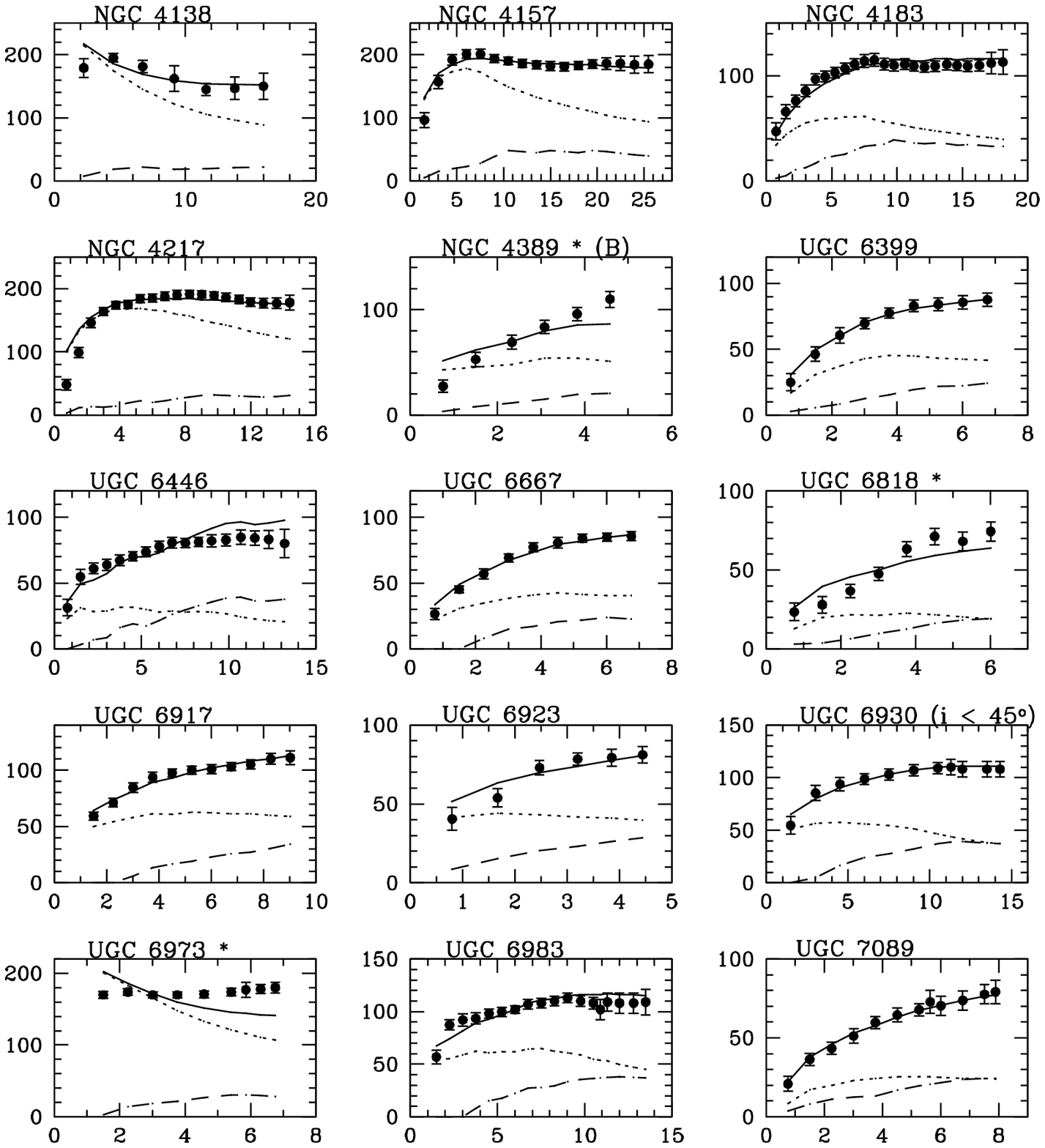}\\
\end{tabular}\par
\caption{Additional MOND rotation curves from Sanders (1996) and de
Blok \& McGaugh (1998) (left) and from Sanders \& Verheijen (1998)
(right). (MOND curves in solid; stellar disc Newtonian curves in
dotted; gas in dot-dash; and stellar bulge in long dashed.)}
\label{fig4}
\end{figure*}
\end{center}

\begin{figure*}[h]
\begin{tabular}{rcl}
\tabularnewline
\includegraphics[width=0.3\columnwidth]{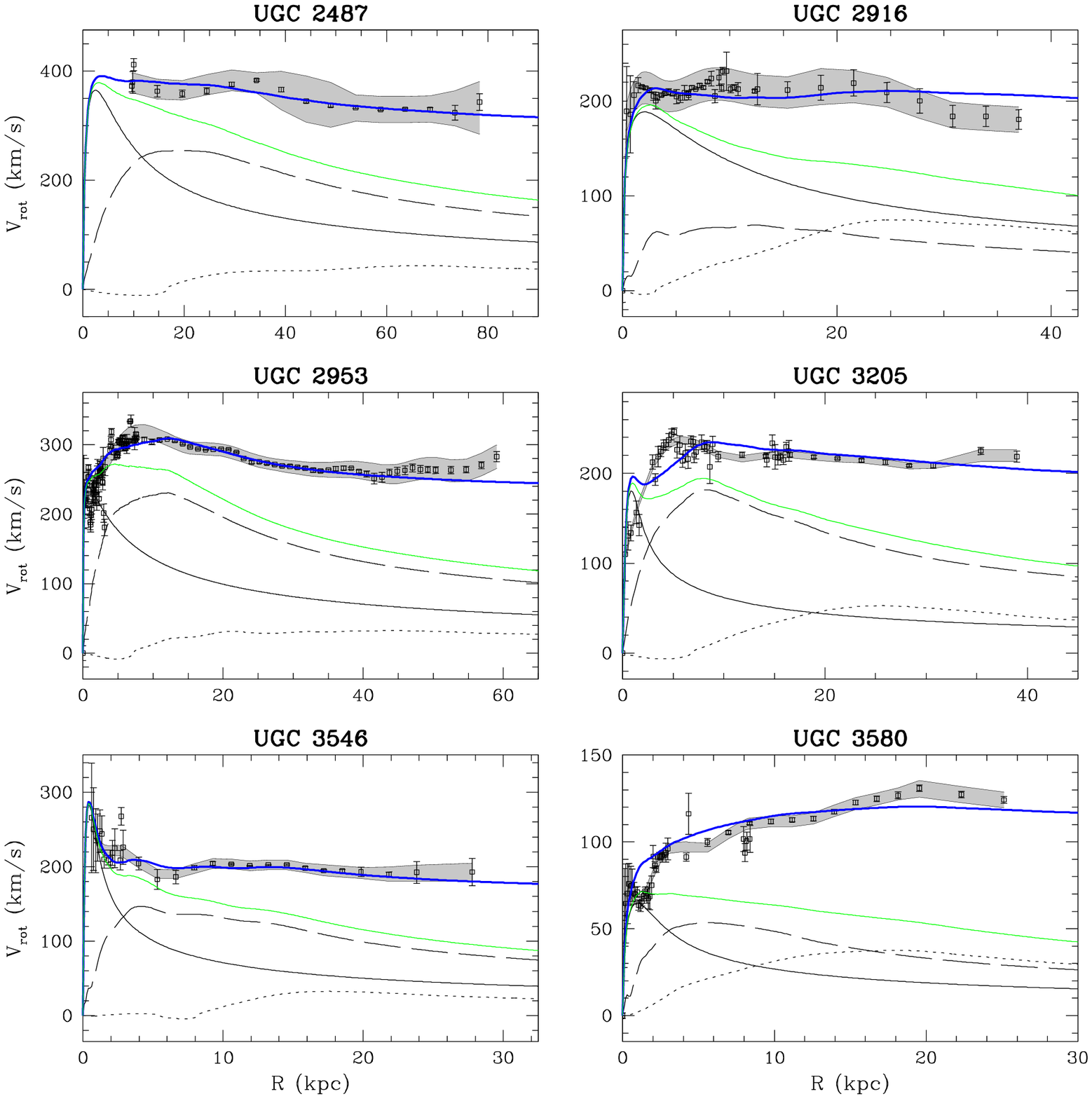} &
\includegraphics[width=0.3\columnwidth]{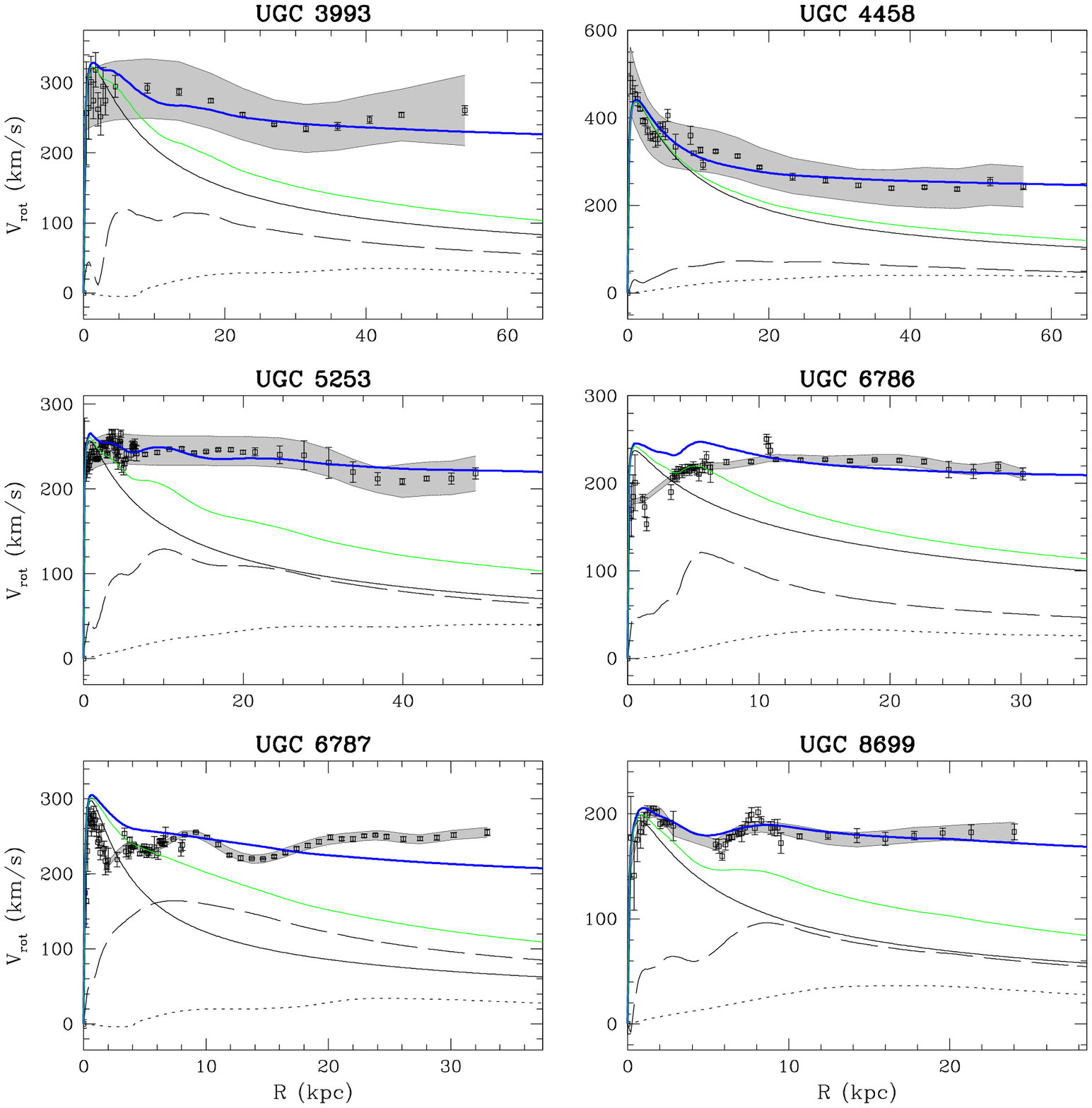} &
\includegraphics[width=0.3\columnwidth]{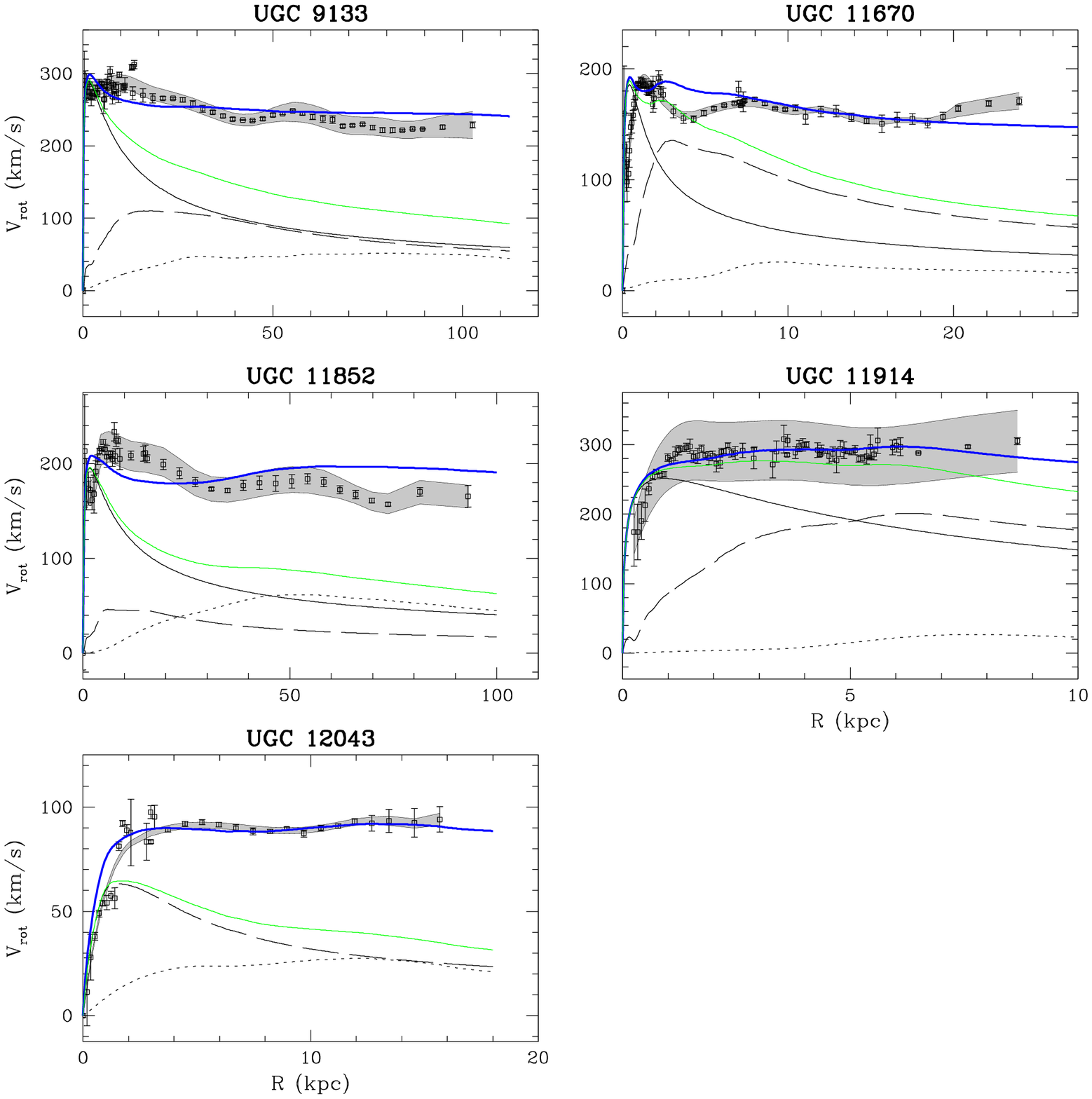}\\
\end{tabular}\par
\caption{Additional MOND rotation curves from Sanders and
Noordermeer (2007).  The grey shaded bands give the allowed range
due to inclination uncertainties. Thin black solid, dashed and
dotted lines give the contributions from stellar bulge, disc and gas
respectively. The thin green (grey) line gives the Newtonian sum of
the individual components and the bold blue (grey) lines gives the
total MOND rotation curve.}\label{fig5}
\end{figure*}

Figure \ref{fig2} shows examples of MOND rotation curve analysis for
three galaxies of very different types. In the center is NGC 1560, a
very low acceleration, gas dominated galaxy, that has a rising RC
within the observed baryons. To the left is NGC 3657, an
intermediate case with similar contributions from the stars and gas.
To the right is NGC 2903, a high acceleration galaxy dominated by
stars, with a declining RC (after the inevitable initial rise). For
NGC 1560, the MOND rotation curve is practically a prediction: since
stellar mass contributes very little the $M/L$ fit parameter gives
hardly any leverage. In addition, since the accelerations are very
small everywhere, the exact form of the interpolating function is
immaterial. The same is true for quite a number of low surface
galaxies of this type. For the other two galaxies, and the many
others like them, the fit $M/L$ parameter does have leverage, but a
very limited one: We can view it, for example, as determined by the
very inner part of the RC, so that the rest of the RC shape and
amplitude becomes an exact, unavoidable prediction of MOND. To boot,
the resulting best fit $M/L$ values are not completely free; they
have to fall in the right ballpark dictated by population synthesis,
as, indeed, they do (e.g. Sanders \& Verheijen 1998). Figs.
\ref{fig3}, \ref{fig4}, \ref{fig5} are mosaics of additional MOND RC
results.

\section{Round systems}
\label{section5} MOND analysis of globular clusters, dwarf
spheroidals, elliptical galaxies, galaxy groups, galaxy cluster,
and even one case of a super cluster have been considered (see
reviews in Sanders \& McGaugh 2002, and in Scarpa 2006). Here I
shall concentrate only on galaxy clusters, which have not yet
fully conformed to MOND's predictions.

\subsection{Cluster dark matter in light of MOND}
 \label{subsection5}
As was realized years ago, MOND does not yet fully account for the
mass discrepancy in galaxy clusters. The first MOND analysis of
clusters (Milgrom 1983c) lowered the deduced $M/L$ values by a large
factor ($\sim 5-10$) compared with Newtonian values. But, it stil
left some clusters with values of a few tens solar units, compared
with several hundreds solar units gotten from Newtonian analysis. At
the time it was wrongly thought that stars exhaust the baryonic
budget of clusters; so $M/L$ was taken to represent the mass
discrepancy. In Milgrom (1983c) I speculated that the x-rays then
known to emanate from clusters may bespeak the presence of large
quantities of hot gas that will remove much of the remaining
discrepancy. This has been largely vindicated by the identification
and weighing of the hot intracluster gas. Reckoning with the gas
reduces the discrepancy, in both MOND and Newtonian dynamics, by a
factor $\sim5-10$, but this is still not enough. Studies based on
gas dynamics and on lensing have helped pinpoint the remaining
discrepancy in MOND, as described by The \& White (1988), Gerbal et
al. (1992), Sanders (1994,1999,2003), Aguirre et al. (2001),
Pointecouteau \& Silk (2005), Angus Famaey \& Buote (2007), and
Takahashi \& Chiba 2007. This remaining discrepancy has to
attributed to yet undetected matter, likely in some baryonic form
(hereafter, CBDM--for cluster baryonic DM). Other suggestions
involve massive neutrinos (Sanders 2003, 2007).
\par
The following rough picture emerges regarding the distribution of
the CBDM: Inside a few hundred kiloparsecs of the cluster center
MOND makes only a small correction; so, most of the discrepancy
observed there must be due to CBDM. The ratio, $\l$, of accumulated
CBDM to visible baryons there is $\sim 10-50$. The results of
Sanders (1999) scaled to $\hubble{75}$ correspond roughly to $\l=2$
at $1~\mpc$. (Pointecouteau \& Silk 2005 claim higher values. But
after correcting for two oversights on their part their results are
consistent with those of Sanders.) The ratio $\l$ decreases
continuously with radius (see, e.g., the small-sample study of
Angus, Famaey \& Buote 2007). At the largest radii analyzed the gas
mass still increases faster than the MOND dynamical mass meaning
that $\l$ is still decreasing there. We can extrapolate to higher
radii and conclude that for the cluster as a whole $\l$ is about 1
or even smaller. McGaugh (2007) reaches a similar conclusion based
on extrapolating the mass-velocity relation from galaxies to
clusters. In summary: in clusters at large the total mass in CBDM is
comparable to that in the baryons already observed, but the CBDM is
rather more centrally concentrated.  So, another factor of two in
the total baryon content of clusters should remove the remaining
discrepancy in MOND. In this connection the historical lesson from
the discovery of the hot gas might be of some value.
\par
Note that the CBDM contributes only little to the total baryonic
budget in the universe. Fukugita \& Peebles (2004) estimate the
total contribution of the hot gas in clusters to $\Omega$ to be
about 0.002. It follows from the above that the contribution of
the CBDM is also only some 5 percent of the nucleosynthesis value.
\par
If the CBDM is made of compact macroscopic objects--as is most
likely-then, when two clusters collide, the CBDM will follow the
galaxies in going through the collision center practically intact,
whereas the gas components of the two clusters coalesce at the
center. Clowe et al. (2006) have recently used weak lensing to map
the mass distribution around a pair of colliding clusters and indeed
found dark mass concentrations coincident with the galaxy
concentrations to the sides of the gas agglomeration near the center
of collision (See, however, Mahdavi et al. 2007 who claim a
conflicting behavior). Such observations are consistent with what we
already know about MOND dynamics of single clusters (Angus et al.
2007).

\par
The presence of CBDM may in the end turn into a blessing as it
might help alleviate the cooling flow puzzle in clusters (Milgrom
2008a).
\par
\begin{figure*}[h]
\begin{tabular}{rll}
\tabularnewline
\includegraphics[width=0.25\columnwidth]{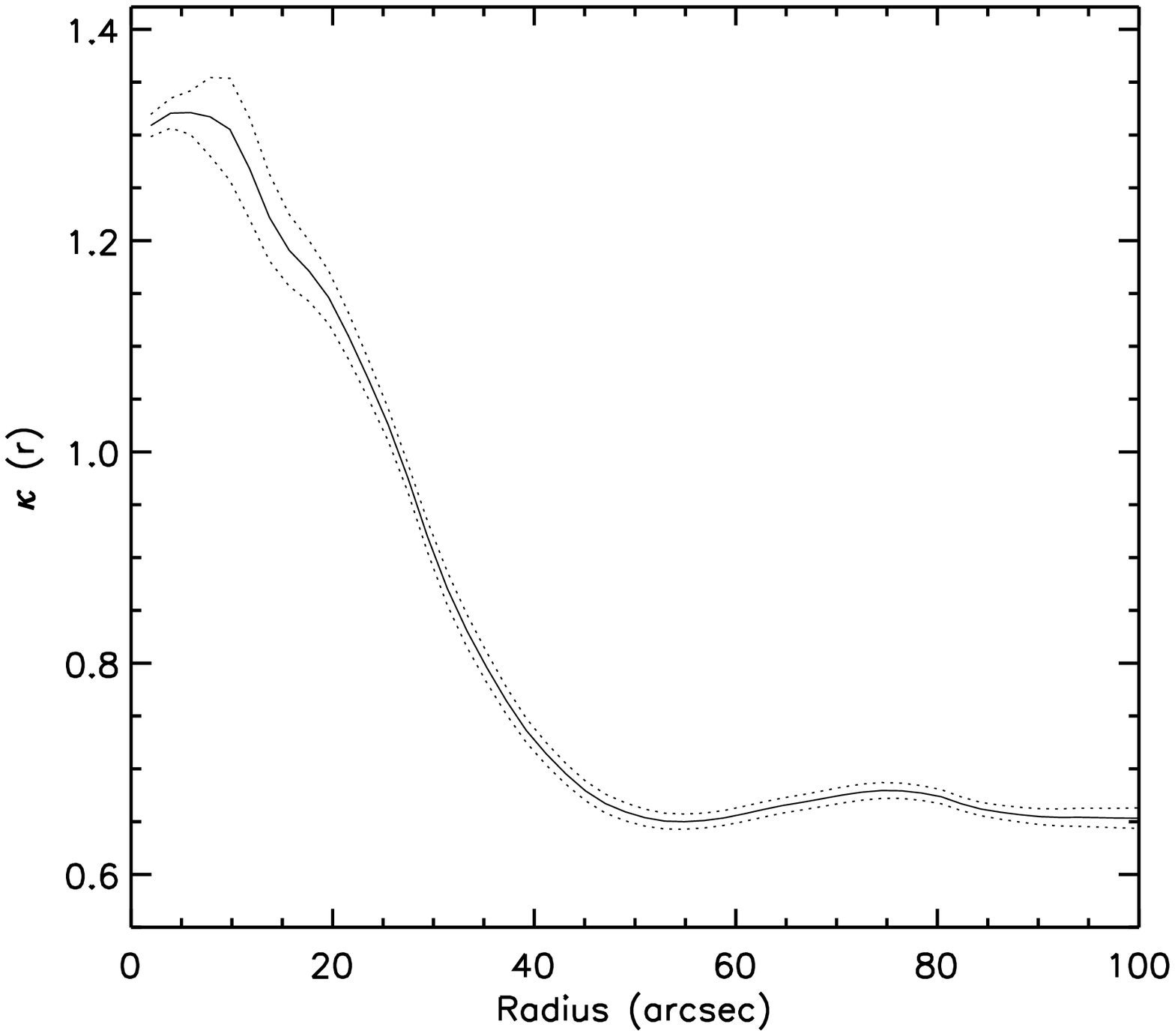}  &
\includegraphics[width=0.3\columnwidth]{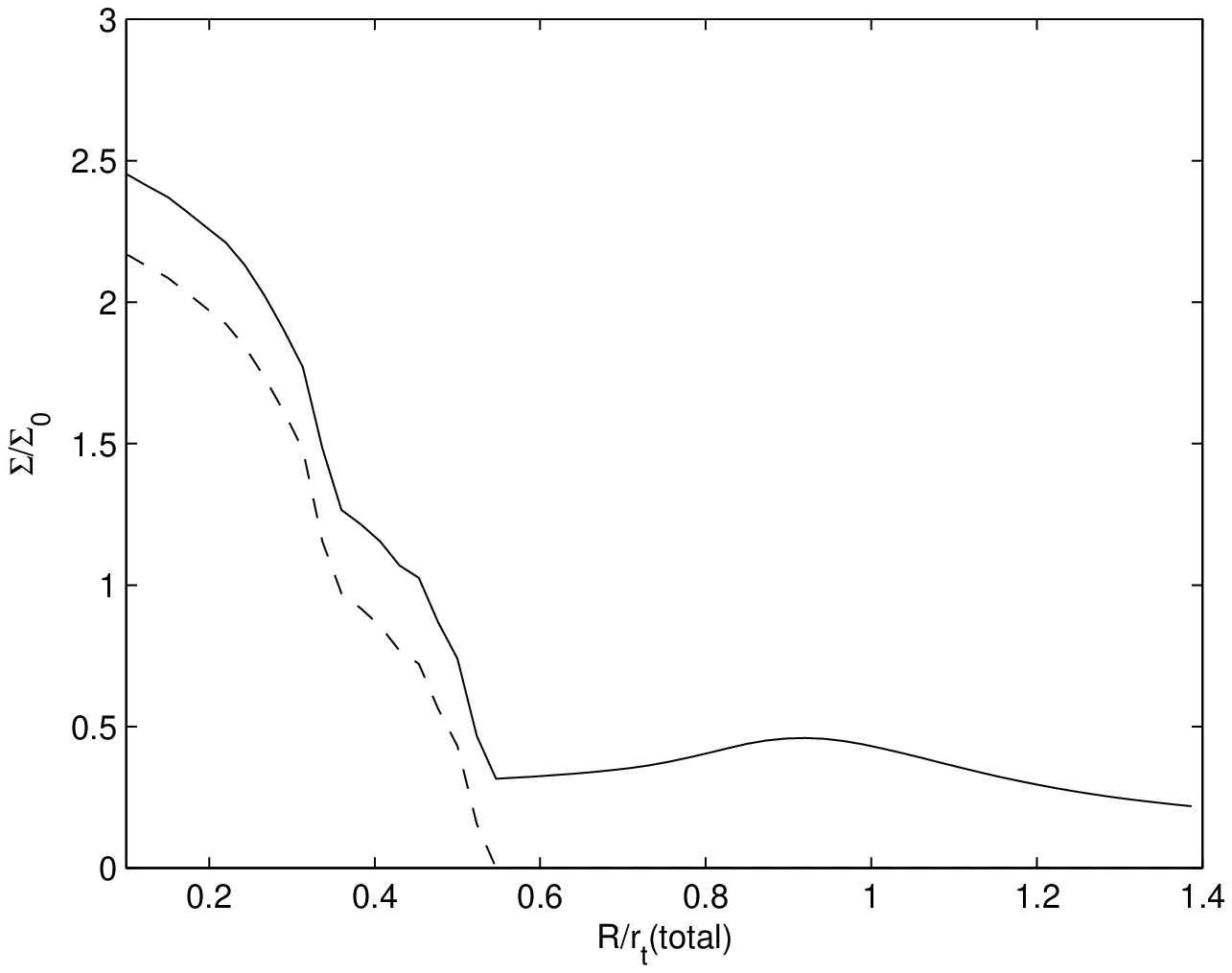}&
\includegraphics[width=0.3\columnwidth]{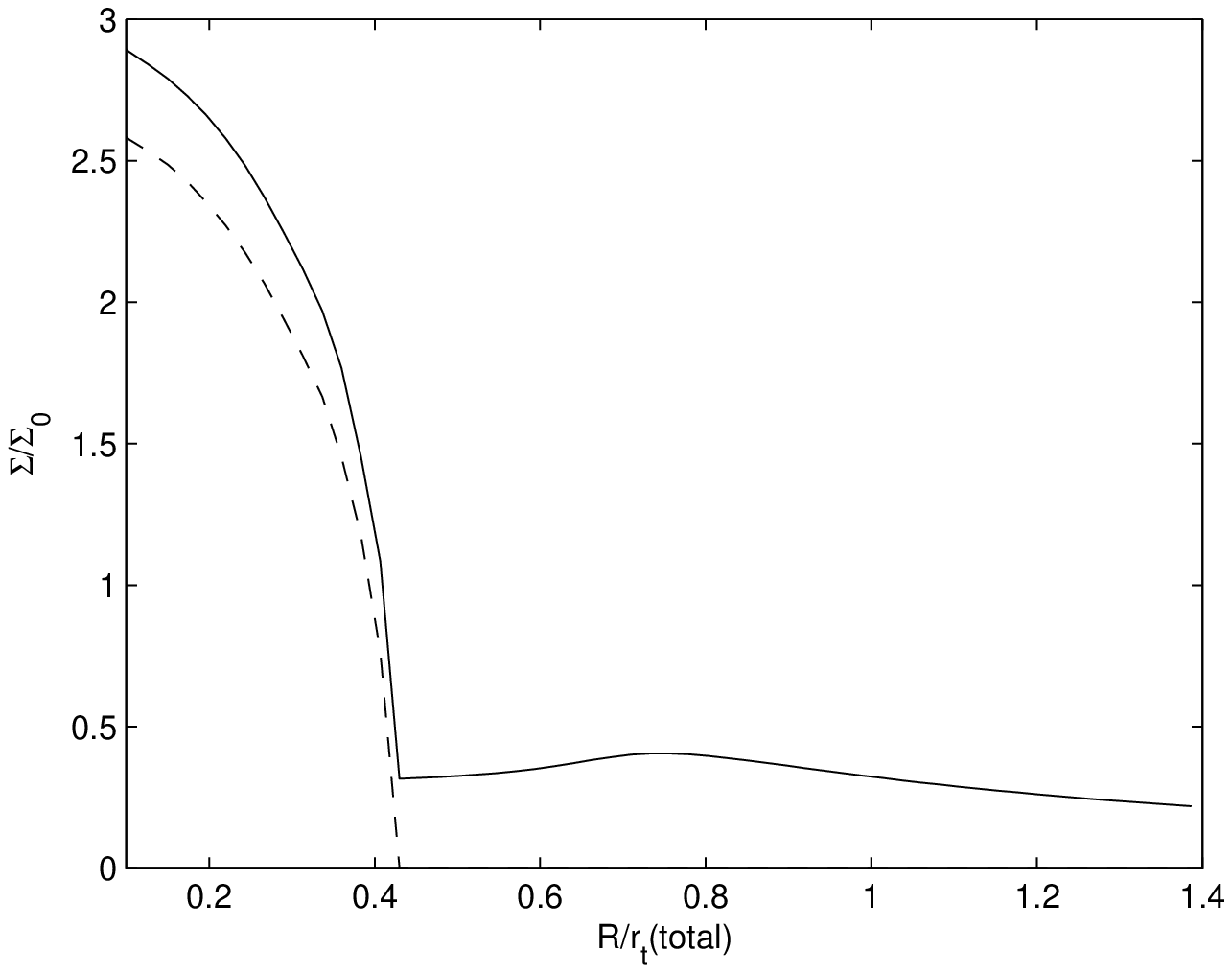}
\tabularnewline \tabularnewline
\end{tabular}\par
\caption{On the left: the deduced surface density distribution
around the cluster Cl 0024+17 using weak lensing (Jee et al. 2007).
The small bump is the alleged ring. The other two frames show the
MOND predictions for very simplistic configurations produced by
masses well contained within the radius of the bump taken from
Milgrom \& Sanders (2008).}
 \label{fig6}
\end{figure*}

Another interesting observation that concerns DM in galaxy clusters
is brought to light by the recent claim by Jee et al. (2007) of a
ring-like structure of DM surrounding the central part of the
cluster Cl 0024+17. This was deduced on the basis of their weak
lensing analysis that shows an enhancement (a bump) in the
distribution of dynamical surface density, as reproduced in the left
panel of our Fig. \ref{fig6}. This was brought up as a potential
difficulty for MOND since, so it was claimed, no corresponding bump
in the distribution of observed baryons is seen. If the presence of
such a ring is confirmed it could be due to CBDM, which we know, in
any case, has to be present (Famaey et al. 2007). However,
interestingly, as shown by Milgrom \& Sanders (2008), such a ring,
with the observed characteristics, could also arise naturally as an
artifact of MOND around the transition radius of the central mass
(see prediction 4 above). A shell of phantom DM is predicted there,
which appears as a projected ring. Verifying the ring as such a MOND
effect would be very exciting as it will constitute a direct image
of the transition region of MOND, analogous to the transition region
of General Relativity as manifested by the formation of a horizon
near $MG/c^2$. Figure \ref{fig6} also shows, in the center and right
panels, the predictions of MOND for the results of weak lensing
around two rather simplistic mass distributions, with some favorable
choices of the interpolating functions.

\section{Solar System}
The acceleration field of the sun is higher than $\az$ within $\sim
8\times 10^3$ astronomical units; so, by and large, MOND affects the
motion of planets and spacecraft in the solar system only in its
very high acceleration limit. In terms of the interpolating
function, MOND predictions for such motion depend on the behavior of
$\mu(x)$ for $x\gg 1$. We have no knowledge of this limit from
galaxy dynamics, which probes $\mu$ to argument values of only a
few. The fact that MOND effects on the planetary motions have not
been observed has been used to constrain the high $x$ behavior of
$\mu(x)$ ever since the advent of MOND; e.g., by Milgrom (1983a),
Sereno \& Jetzer (2006), Iorio (2007), and Bruneton \&
Esposito-Farese (2007). Solar system constraints on relativistic
theories in the high acceleration regime have been considered e.g.,
by Bekenstein (2004), Sanders (1997,2006b), and Bruneton \&
Esposito-Farese (2007). The potential for testing the low
acceleration regime of MOND in special locations in the solar
system, where accelerations almost cancel,  was discussed by
Bekenstein \& Magueijo (2006); earth bound experiments where
proposed by Ignatiev (2007).
\par
An interesting possibility has been raised that the Pioneer anomaly
(reviewed recently by  Nieto \& Anderson 2007) is due to MOND
effects (e.g. discussion in Milgrom 2002a). The fact that the
claimed anomalous acceleration is $a_{anom}\approx 2\pi\az$ is
particularly suggestive. As explained in Milgrom (2002a), if indeed
the anomaly is due to new physics it may point to the option of
modified inertia, rather than modified gravity. This is because
modified gravity predicts similar effects on the planets, contrary
to what is measured (see also Tangen 2007). In modified inertia,
where one modifies the equations of motion of particles, not the
gravitational field, it is possible for particles to suffer very
different MOND corrections at the same position, depending on their
trajectory. So, it might be possible to construct theories that
affect the Pioneer spacecraft on their straight, unbound
trajectories without affecting as much the planets on their
elliptical, bound orbits (see section \ref{modin} below).

\section{$\az$ and its cosmological significance}
\label{section8} The constant $\az$ appears in many of the MOND laws
of galactic motion listed in section \ref{section2}. At first its
value was, indeed,  determined by appealing to some of these (2, 5,
6, and 11) as described in Milgrom (1983b). However, the most
accurate handle on $\az$ comes today from rotation curve analysis.
For example, Milgrom (1988) found $\az\approx 1.3\azun$ from Kent's
reanalysis, and Begeman Broeils \& Sanders 1991 found, with better
data, $\az\approx 1.2\azun$ (both used $\hubble{75}$--the value of
$\az$ depends on the assumed distances to galaxies).
\par
Milgrom (1983a) observed that $2\pi\az\approx cH_0$. And, since we
now know that ``dark energy'' makes up most of the closure density
today, namely $\Omega_\Lambda=\Lambda/3H_0^2\sim1$ we also have
$2\pi \az\approx c(\Lambda/3)^{1/2}$.
\par
With the aid of $\az$, $G$, and $c$ we can construct a length scale
$\ell_0\equiv c^2/\az\approx 10^{29}\cm$, and a mass scale,
$M_0\equiv \mz c^4\approx 6\times 10^{23}\msun$. This is similar to
the emergence, in connection with quantum theory, of the Planck
length and the Planck mass, constructed from $\hbar$, $G$ and $c$.
These set the boundary in the space of phenomena beyond which
combined effects of strong gravity and quantum physics are expected.
 Similarly, $\ell_0$ and $M_0$ tell
us where to expect MOND effects combined with strong gravity.
Because $\ell_0$ is of the order of the Hubble radius $\ell_0\approx
2\pi \ell_H$, with $\ell_H\equiv cH^{-1}_0\approx
c(\Lambda/3)^{-1/2}$, and  $M_0\approx 2\pi M_U$, with $M_U\equiv
c^3G^{-1}(\Lambda/3)^{-1/2}\approx c^3G^{-1}H_0^{-1}$, combined
effects are expected only for the universe at large, and thus there
are no local black holes with surface accelerations in the MOND
regime.
\par
These coincidences may point to a very strong ties between MOND and
cosmology connecting perhaps MOND and the ``dark energy'' effects:
Either one is the result of the other or they are both induced by
the same mechanism.  If the parameter $a_c\equiv
M_U/\ell^2_H=cH_0\approx c(\Lambda/3)^{1/2}$ is somehow felt by
local physics it may not be surprising that dynamics is different
for acceleration above and below this value. It has been long
suspected that local dynamics is strongly influenced by the universe
at large, a-la Mach's Principle, but MOND seems to be the first to
supply a concrete evidence for such a connection. This may turn out
to be the most fundamental implication of MOND, beyond its implied
modification of Newtonian dynamics and General relativity, and
beyond the elimination of DM.
\par
One immediately notes that if the coincidence of $\az$ with
cosmologically significant acceleration parameters is a causal one,
it may hint that $\az$ varies with cosmic time (Milgrom
1989b,2002a). For example, if we always have $\az\sim cH_0$ , or if
$\az$ is related to $\Lambda$ and the latter changes with time, then
$\az$ would follow suit. This is not necessarily the case since
$\az$ could be related to $\Lambda$ with the latter being constant.
Interestingly, variations in $\az$ could induce secular evolution in
galaxies and other galactic systems. For example, the mass-velocity
relations dictate that in the deep MOND regime the velocities in a
system of a given mass should vary like $\az^{1/4}$ if they are in
the deep MOND regime. This would induce changes in radius, so as to
preserve adiabatic invariants, such as possibly $rv$ (Milgrom
1989b). Such variations in $\az$ could also provide an anthropic
mechanism for getting $\Omega_{\Lambda}\sim 1$ at the present time
(Milgrom 1989b, Sanders 1998,2001).
\par
Analyzing the data of Genzel et al. (2006) on the rotation curve of
a galaxy at redshift $z=2.38$, I find that they are consistent with
MOND with the local value of $\az$. However, the large error
margins, and the fact that at the last measured point the galaxy is
only marginally in the MOND regime, still allow appreciable time
variations of $\az$.
\par
MOND as it is formulated at present does not provide a clear-cut
tool for treating cosmology. In my opinion, the understanding of
cosmology in MOND may one day come from the same insight and at the
same time as the understanding of the fundamental basis for MOND.

\section{MOND theories}
\label{section3}
\par
One can build different detailed theories on the basis of the above
premises of MOND. These theories will differ as regards their
detailed predictions, but they are expected to share the core
predictions listed above, and more like them. For example, in the
nonrelativistic regime one can modify the Poisson equation for the
gravitational field, or one can modify the kinetic action of
particles.  Ultimately, one would like to extend the MOND basic
premises to the relativistic regime.
\par
There are two general approaches to building MOND theories: The
first is to construct actions that incorporate the MOND tenets and
thus reproduce MOND phenomenology. (Starting from an action has the
usual advantage that it guarantees the standard conservation laws if
the action has the standard symmetries.) These have, so far, took
$\az$ to be a constant of the theory, and usually employ some form
of an interpolating function between the classical and the MOND
regimes. The other approach is to start from some idea as to the
possible origin of MOND and in this way derive $\az$ and $\mu$ from,
or at least relate them to, more basic concepts. At the moment we
only have some very preliminary attempts in this vein.

\subsection{Non-relativistic action formulations} To demonstrate the sort of
embodiments of the MOND tenets that are possible in the
norelativistic regime we start with the action that describes a
Newtonian system of many gravitating point masses

\beq S=-{1\over 8\pi G}\int\drt~(\gf)^2- \sum_i m_i\f(\vr_i) +\sum_i
m_i\int dt ~v_i^2(t)/2.  \eeqno{iii} The first term is the free
action of the gravitational potential field $\f$, the second is the
interaction of the masses with the field, and the third is the free
(kinetic) action of the particles. We seek to modify this theory so
as to incorporate the MOND tenets.

\subsubsection{Modified gravity}
One possibility is to modify only the free action of the
gravitational field. Perhaps the simplest way to do this is to keep
the Lagrangian as a function of only the first derivatives of the
potential. The most general modification retaining rotational
symmetry is then obtained by replacing $(\gf)^2$ in the first term
by $(\gf)^2 F[({\gf/\az})^2]$. As a result the standard Poisson
equation for the gravitational potential is replaced by
 \beq \div[\mu({|\gf|/\az})\gf]=4\pi G\r, \eeqno{vi}
where $\mu(\sqrt{y})\equiv F(1+dlnF/dlny)$ (Bekenstein \& Milgrom
1984). The MOND axioms require that $F$ goes to 1 ($\mu$ goes to 1)
at large arguments, and $F$ goes to $2y^{1/2}/3$ [$\mu(x)$ goes to
$x$] for small arguments. I would classify such a modification as
``modified gravity'' because it leaves intact the equation of motion
of particles in a given field, and it also affects the dynamics only
in systems where gravity is important. This theory and its
implications for galactic dynamic have been discussed extensively by
Bekenstein \& Milgrom (1984), Milgrom(1984,1986b), Ciotti \& Binney
(2004), and others. It has been used in numerical codes to solve for
the MOND gravitational field in many-body calculations that were
applied to various problems e.g., by Brada \& Milgrom
(1999b,2000a,b), Tiret \& Combes (2007a,b), Nipoti, Londrillo \&
Ciotti (2007a,b), and others.

All the MOND laws listed in section \ref{section2} where shown
explicitly to hold in this theory.

\subsubsection{Modified inertia}
\label{modin}

Another possibility (Milgrom 1994a) is to modify the last term in
the action of eq.(\ref{iii}), replacing it by a general action of
the form
 \beq \sum_i m_i
S_K[\az,\{\vr_i(t)\}].\eeqno{v} Here, $S_k$ is a universal
functional of the particle trajectory $\{\vr_i(t)\}$, independent of
the particle's type. It is possibly non-local, and is a function of
$\az$. This modification leaves the Poisson equation intact but
changes the particle equation of motion (Newton's second law,
$\va=-\gf$) into \beq \vA[\{\vr(t)\},\az]=-\gf, \eeqno{vii} where
$\vA$ is a (possibly non-local) functional of the trajectory, and a
function of $\az$. The dependence on the particle mass is such that
the theory automatically preserves the universality of free fall.

MOND tenet (ii) above dictates that for $\az\rar 0$, $S_K\rar \int
v^2/2$, and $\vA\rar \va$, the acceleration. MOND tenet (iii)
dictates that $S_K[\az,\{\vr_i(t)\}]\rar \az^{-1}s_K[\{\vr_i(t)\}]$,
and so $\vA[\{\vr(t)\},\az]\rar\az^{-1}\vQ[\{\vr(t)\}]$ for $\az\rar
\infty$, where $\vQ$ is a functional of the trajectory with
dimensions of acceleration squared. (By multiplying the action by
$\az^2 G$ and redefining $\az\phi\rar\phi$ we can bring the action
to a form where only the product $m_i G\az$ appears, as required.)
\par
I showed in Milgrom (1994a) that for such a theory to obey the MOND
limits and to be Galilei invariant it has to be non-local, and that
this, in fact, has some advantages. Also note that, again,  the
resulting theory must be nonlinear.

I call such theories modified inertia, as they do not modify the
gravitational field, but modify the equations of motion, and this
for whatever combination of forces is in action in the system,
gravitational or not. As an example, Special relativity would count
here as a (time-local, but non-MOND) modification of inertia. It has
$\vA=d(\gamma\vv)/dt=\gamma[\va+\gamma^2(\vv\cdot\va)\vv/c^2].$

It was also shown in Milgrom (1994a), as a general result for such
theories, that the rotation curves of axisymmetric systems are
simply given by eq.(\ref{alge}), where $\mu(x)$ in this case is
derived from the restriction of the kinetic action to circular
orbits: When a general action functional is restricted to uniform,
circular orbits it reduces to a function of only the radius of the
orbit, $R$, and the rotational speed, $V$. On dimensional grounds, a
MOND action that is normalized to have dimensions of velocity
squared must then reduce to the form $I_{circ}=(1/2)V^2i(V^2/r\az)$.
Then $\mu(x)=i(x)[1+\hat i(x)/2]$, where $\hat i=d ln ~i/d ln ~x$
(Milgrom 1994a). For a given theory $\mu(x)$ is the interpolating
function that applies only to circular orbits in an axisymmetric
potential, but for such orbits it is universal (i.e., unique for the
theory, and independent of the exact nature of the potential field).
This is indeed the equation that has been used in all MOND rotation
curve analyses to date. It follows from this that if the kinetic
action contains terms that vanish for circular orbits, such as terms
that are proportional to $(\vv\cdot\va)^2/\az^2$ (which is not
Galilei invariant), they do not enter $\mu(x)$, and thus do not
affect circular trajectories. They do, however, affect linear motion
as strongly as desired. This is how a modified inertia theory might
be constructed that is consistent with planetary motions, while
explaining the Pioneer anomaly.

\subsubsection{Comparison between modified gravity and modified inertia}
It has to be realized, in the first place, that the division between
the two classes of theories is not always clear, especially so in
the relativistic regime. The free matter action in relativity
contains the gravitational degrees of freedom in addition to those
of matter. This means that modifying the kinetic action will change
the gravitational field equations as well as the equation of motion
of matter. On the other hand, modifying the gravitational field also
modifies the equations of motion. For example, the Brans Dicke
theory, which modifies GR, has two equivalent formulations, one in
which the action for gravity (the Einstein-Hilbert action) is
modified--nominally qualifying as modified gravity--another in which
the matter action is modified (modified inertia). But, at least in
the nonrelativistic (NR) regime the distinction is rather clear and
useful.
\par
We saw in section \ref{section2} that many of the major predictions
of MOND follow from the basic tenets and are thus shared by all MOND
theories. It is thus not so easy, with present day knowledge, to
distinguish even between the two classes of theories. But as there
are differences in the predictions of different theories, so there
are class differences between the predictions of the two theory
types. I mentioned already the fact that in NR modified gravity the
gravitational field of a given source distribution is modified, but
not the equation of motion; so the acceleration of test particles
depends only on their position. In modified inertia the local
acceleration depends also on the trajectory of the particle (as in
Special Relativity). There is still, in such theories a momentum
whose time derivative depends only on position, but this is not the
acceleration.
\par
Another difference is in the exact prediction for rotation curves.
While for modified inertia eq.(\ref{alge}) is exact, it is only
approximate for modified gravity.
\par
Yet another difference is in the definition of the conserved
quantities and adiabatic invariants that emerge from the two
classes: In modified gravity the usual expressions for kinetic
energy and angular momentum hold; not so in modified inertia (a
familiar case in point is, again, Special Relativity).

There is, thus, a potential for distinguishing between the two
classes from the observations.

\subsection{Relativistic formulations}
 A detailed review of relativistic formulations of MOND can be found
in Bekenstein (2006), and in Bruneton \& Esposito-Farese (2007). The
state of the art of this effort is the TeVeS theory (Bekenstein
2004) and its reformulations-generalizations (e.g., Sanders 2005,
Zlosnik Ferreira \& Starkman 2006,2007), which I now describe very
succinctly.

\subsubsection{TeVeS type theories-description}

\begin{itemize}

\item[1.] Gravity in TeVeS (for Tensor, Vector, Scalar)
is described by a metric $ g_{\alpha\beta}$, as in GR, plus a vector
field, ${ U}_\alpha$, and a scalar field $\phi$. In the formulation
of Zlosnik Ferreira \&  Starkman (2006) the scalar is eliminated on
the expense of adding a degree of freedom in the vector.

\item[2.] Matter is coupled to one combination of the fields, dubbed the physical metric

$\tilde g_{\alpha\beta}\equiv e^{-2\phi}(g_{\alpha\beta} +
{U}_\alpha { U}_\beta) - e^{2\phi} { U}_\alpha { U}_\beta$

\item[3.]  $g_{\alpha\beta}$ is governed by the usual Hilbert-Einstein action.

\item[4.] The vector field is governed by a Maxwell-like action, but
is constrained to have a unit length.

\item[5.] The scalar action can be written as $ S_s =-{ 1\over 2Gk\hat k}\int Q\big[\hat k
(g^{\alpha\beta} -{ U}^\alpha {U}^\beta)\phi_{,\alpha}\phi_{,\beta}
\big](-g)^{1/2} d^4 x. $

\item[6.] There are three constants appearing in this formulation: $k, ~\hat k$, and
$K$,  and one free function $Q(x)$. The last begets the
interpolating function of MOND in the nonrelativistic limit.
\end{itemize}

\subsubsection{TeVeS type theories-results}

\begin{itemize}

\item[1.] For nonrelativistic Galactic systems it reproduces the nonrelativistic MOND phenomenology
yielding for this case eq.(\ref{vi}) with $\az\propto k\hat
k^{-1/2}$.
\item[2.] Lensing in weak fields ($\f\ll c^2$): TeVeS gives lensing according to the standard GR
formula  but with the MOND potential derived from eq.(\ref{vi}).

\item[3.] Structure formation: Preliminary work is described in Sanders (2005),  Dodelson \& Liguori
(2006), Skordis (2006), and Skordis et al. (2006).
 \vspace{0.4cm}
\item[4.] CMB: preliminary work: TeVeS has the potential to mimic aspects of
cosmological DM (Skordis et al. 2006).

\item[5.] Desiderata: $\az$ and the interpolating function are still
put in by hand.

\end{itemize}

\subsection{Effective theories}

The coincidence of $\az$ with cosmic acceleration parameters hints
at the possibility that MOND is an effective, or emergent, theory.
This would mean that MOND might emerge as an approximate consequence
of some deeper physical theory. In such a scheme $\az$ might turn
out not to be a fundamental constant of the underlying theory. This
is similar to the case of the free fall acceleration on earth, which
appears as a constant of nature when we deal with dynamics near the
surface of the earth, but has no significance in the underlying
gravitation theory. Likewise, the interpolating function, or its
equivalents, that appear in MOND theories should be derivable from
the underlying theory. The role of this function in MOND is similar
to that of the black body function in quantum mechanics, or to that
of the Lorentz factor in Special Relativity. They too interpolate
between the appropriate classical limit and the modified regime,
with the fundamental constant of the theory setting the boundary.
They too where introduced first on phenomenological grounds, but
were then derived from a basic underlying theory.
\par
For example, MOND could result  from the imposition of a new
symmetry generalizing Lorentz invariance (Milgrom 2005) in a way
that connects cosmology with local physics. There are several ideas
for possible underlying schemes for MOND. All are at the moment
rather preliminary. I list three of these below.

\subsubsection {Vacuum effects}

Applying MOND as it is now formulated requires knowledge of some
inertial frame with respect to which absolute accelerations can be
measured: Because MOND is nonlinear we need to substitute in its
equations a value of the absolute acceleration. One usually assumes
that this is the local rest frame of galaxies--the cosmologically
comoving frame (e.g., in applications of the external field effect).
But what could be the physics underlying the choice of such a frame?
Milgrom (1999) pointed out that the quantum vacuum might provide it.
The vacuum constitutes a physical inertial frame in the sense that
any observer with some internal structure can detect its own non
inertial motion with respect to it. This can be done via the Unruh
effect by which a non inertial observer in an otherwise empty
Minkowski vacuum detects Unruh radiation that depends in a
complicated way on the observer's world-line. For the simplest,
nontrivial case of an observer with a constant acceleration, $a$,
the radiation is thermal with a temperature  $T=\alpha a, ~~~~
\alpha\equiv \hbar/2\pi kc$. This, or something like it, could be
the sensor that tells a system that it is being accelerated, and
thus be the marker for inertia. An analog effect takes place even
for an inertial observer in an expanding universe. Again, in the
simplest case of a de Sitter universe an inertial observer finds
itself immersed in thermal radiation with $T=\alpha
c(\Lambda/3)^{1/2}$, where $\Lambda$ is the appropriate cosmological
constant defining the curvature of space time (the Gibbons-Hawking
effect). This may somehow be felt by bodies and imprint an effect of
cosmology in local dynamics, as is hinted by MOND. An observer with
a constant acceleration world line in a de Sitter universe also
senses thermal radiation with temperature
$T=\alpha(a^2+c^2\Lambda/3)^{1/2}$ (Narnhofer, Peter, \& Thirring
1996, Deser \& Levin 1997). The temperature difference between such
an observer and one that is inertial is then
 \beq\Delta T=\alpha[(a^2+c^2\Lambda/3)^{1/2}-c(\Lambda/3)^{1/2}]=\alpha
a\mu(a/\hat\az),\eeqno{huio1}
 with $\mu(x)=[(1+4x^2)^{1/2}-1]/2x$,
and $\hat\az=2c(\Lambda/3)^{1/2}$.  The two limits of this
temperature difference for high and low accelerations are:
 \beq
\Delta T\propto\left\{\begin{array}{l@{\quad:\quad} l}a & a\gg
\hat\az\\a^2/\hat\az & a\ll \hat\az\end{array}\right..\eeqno{nuj}
This is very much reminiscent of what is required for MOND inertia,
with the added bonus of a relation between $\hat\az$ and the
cosmological parameter $\Lambda$. Applied to the Pioneer spacecraft,
using the measured value of the cosmological constant, such an
expression for inertia gives $a=MG/r^2+\hat\az/2$, with
$\hat\az/2\approx 6\times 10^{-8}~{\rm cm~ s}^{-2}$, compared with
the measured value of the anomaly $a_{an}\approx 8\times
10^{-8}~{\rm cm~ s}^{-2}$. Note that we do not live exactly in a de
Sitter universe and that the motion of the Pioneer spacecraft is not
exactly one of constant acceleration, for which situation the above
expression was derived.
\par
Note also that for modified inertia, as would be the case here, we
cannot, without a theory, learn much about circular orbits from
observations of linear ones like those of the Pioneer
spacecraft--with acceleration parallel to the velocity. In
particular the effective interpolating function applicable to
rotation curves need not be the same as that for linear trajectories
that we find here. Also, the fact that $\hat\az$ here isn't exactly
$\az$ we find for rotation curves isn't necessarily meaningful. This
too would just point to a different effective $\mu(x)$ that behaves
as $\lambda x$ at small x (with $\lambda\approx .1$), not as $x$ as
in the case of circular orbits. All this should be expected: if the
Pioneer anomaly is at all a MOND effect there will have to be a very
different effect on quasi circular orbits since, as I mentioned
above, the anomaly is not detected for the planets.

\subsubsection {Membrane models}
Equation (\ref{vi}) is identical in form to the equation that
determines the shape of a membrane that tries to minimize its area
(or it volume, in higher dimensions). The potential $\f$ then stands
for the height of this membrane above some reference plane. (See,
e.g., Milgrom 2002c for a discussion of this and other physical
systems governed by an equation of the same form.) However, the
resulting $\mu$ function is not that of MOND. Milgrom (2002a)
discussed a scenario in which our 3-space is a 3-D membrane moving
in a 4-D space. In this last there is a preferred direction, in
which the membrane is being accelerated. The position along this
direction is perceived on the membrane as the gravitational
potential (all this is a nonrelativistic description in need of a
relativistic extension). The energy function of the membrane is not
simply its area but one that depends also on the orientation of the
membrane with respect to the preferred direction.

\subsubsection {Polarizable medium}
Equation(\ref{vi}) is also identical in form to the equation for the
electrostatic potential in a nonlinear, dielectric medium in which
the dielectric constant is a function of the field strength (e.g.,
Milgrom 2002c). This has lead Blanchet (2007a,b, see also this
volume) to propose a physical theory for MOND based on a
gravitationally polarizable medium.

\section{Significance for DM}
\label{section9}

It is sometimes claimed by DM advocates that the successes of MOND
will one day be understood in terms of DM, meaning that MOND
somehow summarize how DM acts. This cries for undeceiving: Can the
myriad observations on the distribution of ``DM'' be all gotten
from the baryon distribution alone through a very simple formula
involving one universal parameter? Can the ubiquitous appearance
of the constant $\az$ in seemingly independent galactic phenomena
emerge somehow from the DM paradigm?
\par
The nature and origin of mass discrepancies in galactic systems
differ greatly in MOND and in the Newtonian-dynamics-plus-DM
paradigms. In MOND, these discrepancies are not real, they are
artifacts of adhering to Newtonian dynamics instead of MOND. As a
result, MOND predicts them uniquely from the presently observed
(baryonic) mass distribution. As we saw, the pattern of these
discrepancies is predicted, and is observed, to follow a large
number of well defined relations. For MOND to be some summarizing DM
formula we will have to conclude that the distribution of baryons
fully determines that of the DM. However, in the DM paradigm the
expected distributions of the two components are strongly dependent
on details of the particular history of a system: The two types of
matter  are very different: baryons are dissipative and  strongly
interacting with photons and magnetic fields; CDM supposedly is
neither. Along the haphazard history of a galactic system they are
then subject to different influences. The formation process and the
ensuing unknown and {\it unknowable} history of mergers,
cannibalism, gas accretion, ejection of baryons by supernovae and
ram pressure, energy loss by dissipation, interaction with magnetic
fields, etc., all affect baryons and DM differently. These processes
are expected to produce haphazard relative amounts and distributions
of the two components in the system. The fact that baryons and DM
are well separated today, and that baryons form discs while CDM does
not are obvious results of such differentiation. Another strong and
direct evidence comes from the recent realization that the ratio of
baryons to required DM in present day galaxies is smaller by an
order of magnitude, typically, than the cosmic value, with which
protogalaxies presumably  started their life . This evidence comes,
for example, from probing large galactic radii with weak lensing ;
e.g. by Kleinheinrich et al. (2004), Mandelbaum et al. (2005), and
by Parker et al. (2007) (see also McGaugh 2007 for evidence based on
small radius data with CDM modeling). This means that in the DM
paradigm, galaxies should have somehow lost most of their baryons
($\sim 90\%$) during their history. Even for galaxy clusters there
is now some tentative evidence that the observed baryon fraction is
a few tens of percents smaller than the cosmic value (Afshordi et
al. 2007 and references therein).
\par
\begin{figure*}[h]
\begin{tabular}{rcl}
\tabularnewline
\includegraphics[width=0.3\columnwidth]{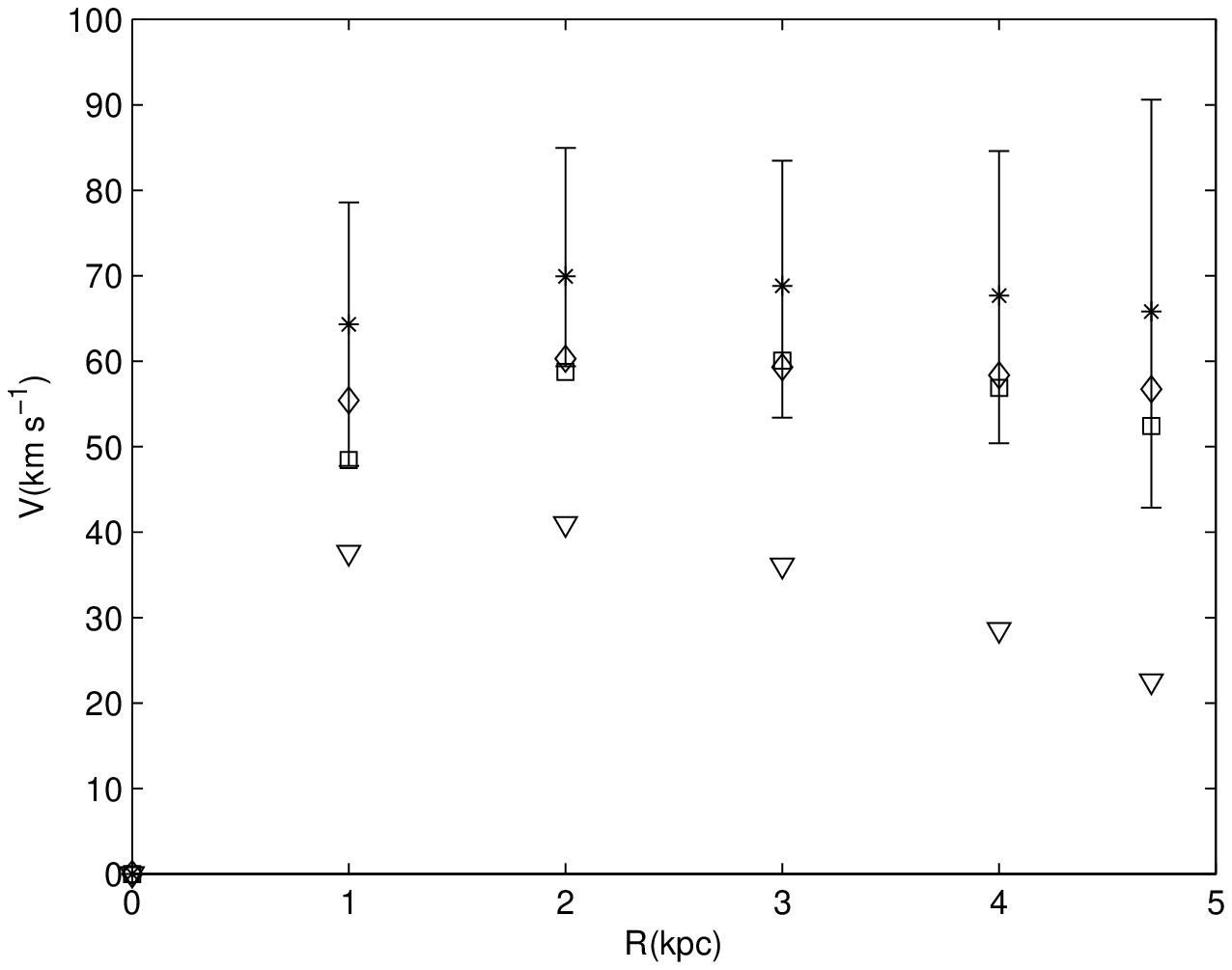} &
\includegraphics[width=0.3\columnwidth]{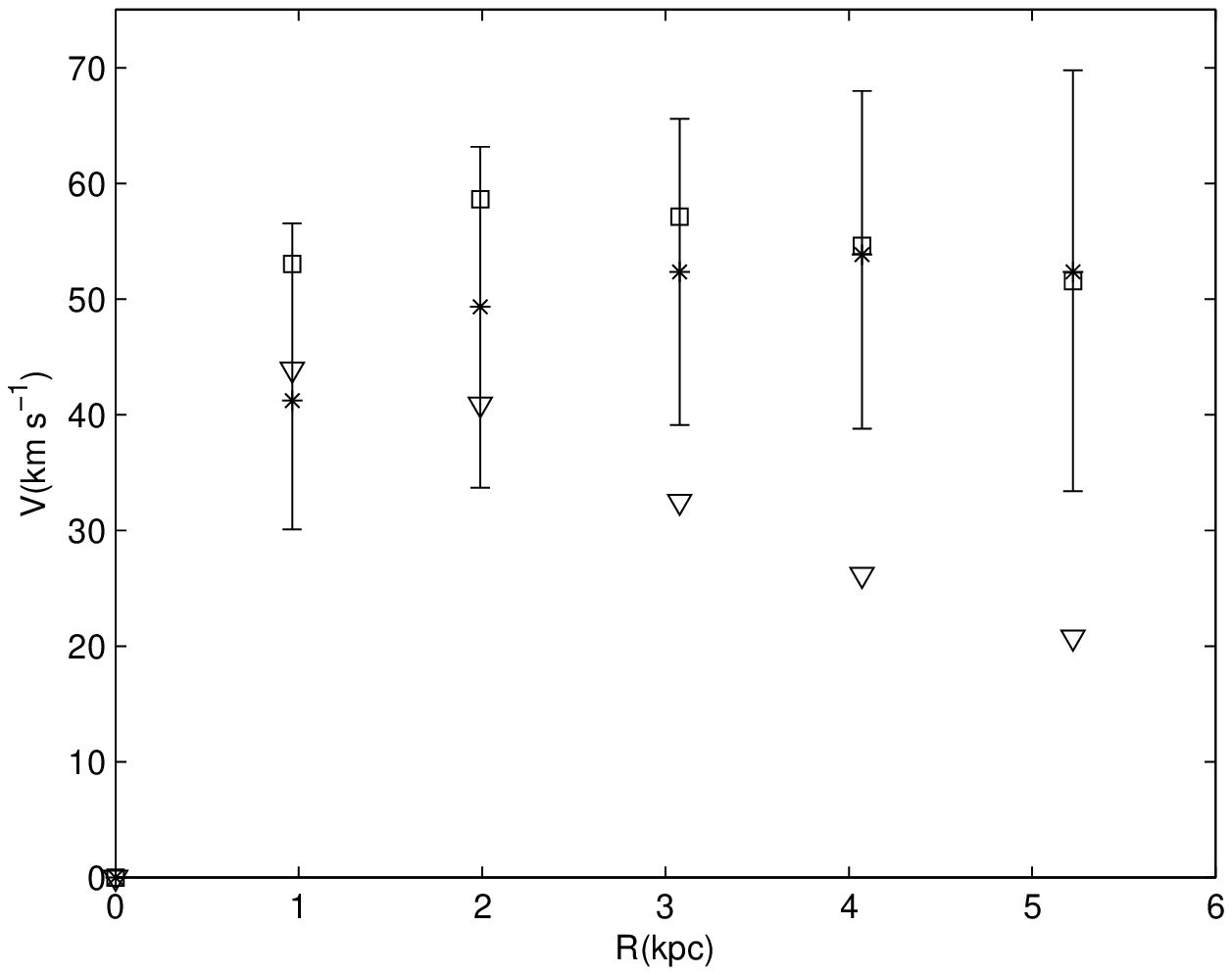} &
\includegraphics[width=0.3\columnwidth]{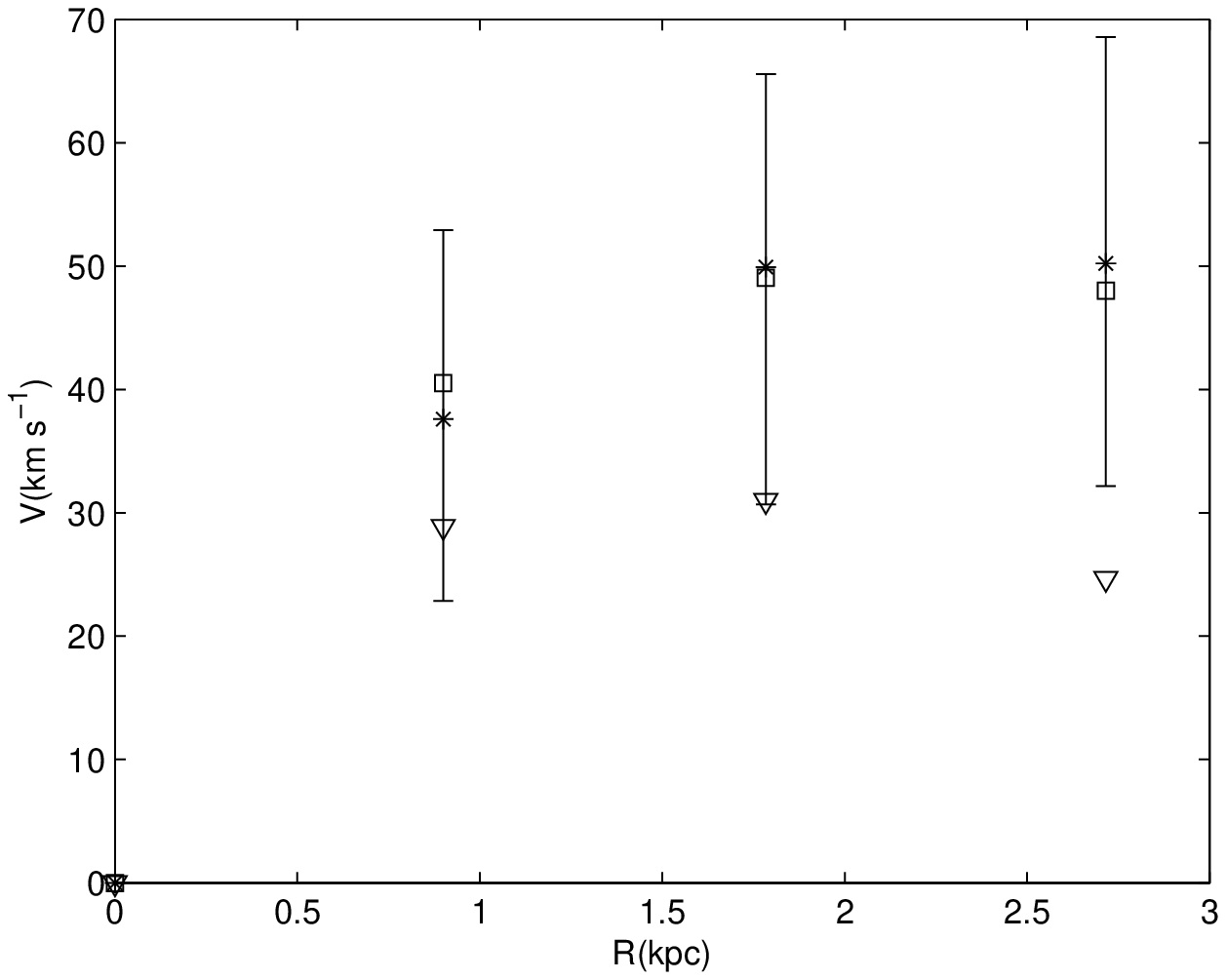}\\
\end{tabular}\par
\caption{The rotation curves for the three debris dwarfs from
Bournaud et al. (2007): NGC 5291N (left), 5291S (center), and 5291SW
(right). The measured velocities, for the nominal inclination of
$45\deg$, are marked by stars and are shown with their error bars.
The calculated Newtonian velocities are marked by inverted
triangles. The predicted MOND velocities are marked by squares. Also
shown, for NGC 5291N only, the measured velocities for an assumed
inclination of $i=55\deg$ (diamonds); from Milgrom
(2007).}\label{fig7}
\end{figure*}

Another poignant example of the large variety of baryon vs. DM
properties expected in the DM paradigm is brought into focus by the
recent observation of large mass discrepancies in three tidal debris
dwarf galaxies (Bournaud et al. 2007). This case is doubly
interesting because it is also one where CDM and MOND predictions
differ greatly. In light of the  specific formation scenario of such
dwarfs, CDM predicts very small amount of DM in them (see discussion
of Bournaud et al.). This is in contrast with what is expected in
primordial dwarfs for which large DM to baryon fractions are
predicted. Since both types of dwarfs are low acceleration systems,
MOND predict large mass discrepancies in both. The dwarfs analyzed
by Bournaud et al. (2007) do show substantial mass discrepancies as
predicted by MOND as shown in Fig. \ref{fig7} (Milgrom 2007, Gentile
et al. 2007b), and contrary to what CDM predicts.

\par
In a DM interpretation of MOND one will then conclude that the
haphazard and small amount of leftover baryons in galaxies determine
so many of the properties of the dominant DM halo, with such
accuracy as evidenced by MOND. I deem it highly inconceivable that
DM will ever reproduce the predictions of MOND and the relations it
predicts for individual systems. Achieving this within the complex
scenario of galaxy formation would be akin to predicting the
detailed properties of a planetary system--planet masses, radii, and
orbits--from knowledge of only the present day central star. When we
find a tight relation between system properties in astronomy--such
as the zero-age main sequence for non-rotating, nonmagnetic stars of
a given composition--it results inescapably from laws of physics. We
cannot conceive of an incipient star (non-rotating, non-magnetic,
etc.) that does not sit on the main sequence; we cannot conceive of
a galaxy in MOND that does not satisfy the MOND laws; but we can
easily conceive of such a galaxy in the Newtonian dynamics plus DM
paradigm.
\par
Indeed, to my knowledge none of the MOND predictions listed in
section \ref{section2} has been shown to follow in the DM paradigm.
(I discount the attempt by Kaplinghat and Turner 2002 to explain
prediction 3, for the reasons I gave in Milgrom 2002b.)
\par
For example, it is being claimed that LCDM predicts a Tully-Fisher
relation not unlike what is observed. This, however, is not really
true. What LCDM predicts is a relation between the total halo mass
and its maximum circular-orbit velocity. In earlier times it could
be assumed with impunity that the ratio of total baryons to total DM
mass in galaxies is universal (and equal to the cosmological value).
This would have given a relation between the baryon mass and the
rotational velocity for the halo. However, the assumption of a
universal baryon fraction is now known not to follow in the CDM
picture, and, as we saw, this fraction is typically much smaller
than the putative cosmological value. Not only isn't the general
correlation predicted, but the processes that caused only a small
fraction of the original baryons to show up in present day galaxies
are likely to have produced large scattering in this ratio, and
hence in the predicted baryon-mass-velocity relation, unlike what is
observed. (Simulations that include baryons are very ad hoc, and
introduce many assumptions by hand. Clearly, true simulations of
baryon behavior in this context are far beyond the offing.)
\par
Regarding the ability of DM models to fit rotation curves of disc
galaxies: Apart from the well known lingering problem of fitting the
inner parts of galactic rotation curves (the ``cusp problem''),
halos predicted by LCDM (with, e.g., NFW profiles) can, by and
large, fit the observed curves, but only if one leaves free their
mass and size parameters. Such fits then involve three parameters:
the stellar mass-to-light ratio (the only parameter used in MOND
fits) plus the two structural parameters for the halo, which afford
great freedom in reconstructing rotation curves. However, in LCDM
simulations the two structural parameters are not free, but come out
strongly correlated. If one actually uses the halo profiles
predicted by LCDM, with that correlation enforced, the rotation
curve fits are bad, even though they still have the freedom of an
additional halo parameter vis-a-vis MOND (e.g., McGaugh et al. 2007,
Gentile Tonini \& Salucci 2007, Gentile et al. 2007a).
\par
Dark matter advocates themselves invoke ``baryon-less'' galaxies
when they want to explain the ``missing satellite'' problem for CDM.
But this would then contradict the existence of strict relations
between baryons and DM, which they would have to invoke to explain
MOND predictions within the DM paradigm. The blanket is too short to
cover both ends.
\par
To recapitulate, the confirmations of the MOND predictions--e.g.,
those listed in section \ref{section2} and those concerning the full
rotation curves--argue against the Newtonian-dynamics-plus-DM
paradigm in two ways. First, since they supports MOND as a competing
paradigm. Second, because in themselves, and without reference to
MOND, these regularities point to a strong baryon-DM connection in
which Baryons determine completely the distribution of DM in a many
well defined and independent ways, and object by object. This would
clearly fly in the face of the expectations from DM in which the
relation between baryons and DM is haphazard and strongly dependent
on the history of each object.

\section*{Acknowledgements}

I thank Jacob Bekenstein and Bob Sanders for helpful comments on the
manuscript and, even more, for the many exchanges of ideas on MOND
we have all had over the years. This research was supported by a
center of excellence grant from the Israel Science Foundation

\bibliographystyle{elsart-harv}

\clearpage
\end{document}